\documentclass{aa}

\usepackage{txfonts}
\usepackage{xcolor}
\usepackage{physics}
\usepackage{graphicx}
\usepackage{booktabs}
\usepackage{multirow}

\definecolor{softblue}{HTML}{30688E}
\definecolor{softgreen}{HTML}{35B778}

\newcommand{\np}{n_{\mathrm{p}}}
\newcommand{\nel}{n_{\mathrm{e}}}
\newcommand{\ngam}{n_{\gamma}}
\newcommand{\Yng}{Y_{\nu \gamma}}

\usepackage[colorlinks=true, linkcolor=softblue, citecolor=softblue, urlcolor=softgreen]{hyperref}

\bibpunct{(}{)}{;}{a}{}{,}

\graphicspath{{./}{figures/}}
\defcitealias{IceCube:2018dnn}{IC18}

\begin{document}

% \linenumbers

\title{Assessing coincident neutrino detections using population
  models}

\author{F. Capel\inst{\ref{mpp},\ref{tum}} \and J. M.
  Burgess\inst{\ref{mpe}} \and
  D. J. Mortlock\inst{\ref{imp_astro},\ref{imp_stats},\ref{okc}} \and
  P. Padovani\inst{\ref{eso},\ref{inaf}}}

\institute{Max Planck Institute for Physics, Föhringer Ring 6 D-80805
  Munich, Germany \email{capel@mpp.mpg.de} \label{mpp} \and Technical
  University of Munich \& ORIGINS Excellence Cluster,
  Boltzmannstra{\ss}e 2, D-85748 Garching, Germany \label{tum} \and
  Max Planck Institute for Extraterrestrial Physics,
  Giessenbachstra{\ss}e 1, D-85748 Garching, Germany \label{mpe} \and
  Astrophysics Group, Imperial College London, Blackett Laboratory,
  Prince Consort Road, SW7 2AZ London, UK \label{imp_astro} \and
  Statistics Section, Department of Mathematics, Imperial College
  London, SW7 2AZ London, UK \label{imp_stats} \and The Oskar Klein
  Centre, Department of Astronomy, Stockholm University, AlbaNova,
  SE-10691 Stockholm, Sweden \label{okc} \and European Southern
  Observatory, Karl-Schwarzschild-Str. 2, D-85748 Garching,
  Germany \label{eso} \and Associated to INAF - Osservatorio di
  Astrofisica e Scienza dello Spazio, via Piero Gobetti 93/3, I-40129
  Bologna, Italy \label{inaf} }

\date{}

\abstract{Several tentative associations between high-energy neutrinos
  and astrophysical sources have been recently reported, but a
  conclusive identification of these potential neutrino emitters
  remains challenging. We explore the use of Monte Carlo simulations
  of source populations to gain deeper insight into the physical
  implications of proposed individual source--neutrino
  associations. In particular, we focus on the IC170922A--TXS~0506+056
  observation. Assuming a null model, we find a 7.6\% chance of
  mistakenly identifying coincidences between $\gamma$-ray flares from
  blazars and neutrino alerts in 10-year surveys. We confirm that a
  blazar--neutrino connection based on the $\gamma$-ray flux is
  required to find a low chance coincidence probability and,
  therefore, a significant IC170922A--TXS~0506+056 association. We
  then assume this blazar--neutrino connection for the whole
  population and find that the ratio of neutrino to $\gamma$-ray
  fluxes must be $\lesssim 10^{-2}$ in order not to overproduce the
  total number of neutrino alerts seen by IceCube. For the
  IC170922A--TXS~0506+056 association to make sense, we must either
  accept this low flux ratio or suppose that only some rare
  sub-population of blazars is capable of high-energy neutrino
  production. For example, if we consider neutrino production only in
  blazar flares, we expect the flux ratio of between $10^{-3}$ and
  $10^{-1}$ to be consistent with a single coincident observation of a
  neutrino alert and flaring $\gamma$-ray blazar. These constraints
  should be interpreted in the context of the likelihood models used
  to find the IC170922A--TXS~0506+056 association, which assumes a
  fixed power-law neutrino spectrum of $E^{-2.13}$ for all blazars.}

\keywords{neutrinos -- astroparticle physics -- methods: data
  analysis}

\maketitle

\section{Introduction}
\label{sec:intro}

Multiple observations of the same sources, across different
wavelengths or probes, can give valuable insight into the underlying
physical processes that are responsible for this emission, as
demonstrated by multi-messenger astronomy. However, as more data are
collected with new detectors and larger surveys, and as we search more
thoroughly with targeted follow-up programmes, it becomes increasingly
likely for us to observe phenomena that may appear to be connected,
but are in fact just coincident by chance.

Of course, there are cases where observations are obviously connected,
such as that of GW170817 and GRB 170817A \citep{Abbott:2017hz}, and
cases where they are obviously disconnected. What drives our initial
judgement is typically the spatial and temporal relationship of the
different signals, compared to what is expected from theoretical
considerations. In between these two extremes, it is not uncommon to
find potential connections that remain inconclusive due to poor signal
localisation or uncertain temporal connection (see
e.g. \citealt{Kadler:2016cx}, \citealt{Graham:2020iy} and
\citealt{Ajello:2021ks} for recent examples).

This latter scenario is also particularly relevant for the ongoing
search for astrophysical neutrino sources. Proposed associations are
uncertain, and we expect signals to appear weak compared to known
backgrounds in the data. A joint collaboration of several instrument
teams, including the IceCube and Fermi-LAT collaborations, have
reported the association of a $\sim$~290~TeV neutrino and the blazar
TXS~0506+056 (\citealt{IceCube:2018dnn}, hereafter
\citetalias{IceCube:2018dnn}). The neutrino and source are
directionally consistent on the sky, within uncertainties; the
neutrino has a $56.5\%$ probability of being astrophysical and it is
seen to arrive during a 6-month active period of the blazar, in which
the $\gamma$-ray activity is increased. The resulting significance is
found to be at the $3\sigma$ level. If true, this association has
profound implications for our understanding of hadronic acceleration
in blazars and so it is pertinent to develop deeper and complementary
analyses using available information to try to resolve these open
questions.

One way to evaluate associations is to utilise more of the available
data by developing the statistical methods that are used to study
individual event--source associations. Several recent efforts in this
direction are based on Bayesian frameworks that can be extended to
include more information on the event--source connection (see
e.g. \citealt{Ashton:2018gq}, \citealt{Capel:2019hn},
\citealt{Bartos:2019hj}, \citealt{Veske:2021nb}). Ideally, such
approaches would also involve the information gained from modelling
the multi-wavelength spectra of these objects to determine if neutrino
emission makes sense in the context of possible physical models (see
\citealt{Boettcher:2019jf, Gasparyan:2022} for a recent review).
  
It is also important to consider the implications of potential
individual associations in the context of the relevant astrophysical
source populations. All sources of interest belong to some class of
sources with similar properties and so the two are inextricably
linked. General constraints on the density and effective luminosity of
an unknown source population can be derived by requiring it to be able
to produce the total astrophysical neutrino flux seen by IceCube,
without containing individual sources that would have been detected by
previous point source searches of the integrated data
\citep{Murase:2016ql, Capel:2020cj}. However, to study these proposed
associations in a meaningful way, a more specific modelling of the
population and its multi-messenger connections is necessary.

In this work, we present a conceptually straightforward Monte Carlo
simulation strategy for assessing the validity of proposed
associations in the context of the relevant source populations. Here,
we take the case of the blazar--neutrino association described in
\citetalias{IceCube:2018dnn} as an interesting case study. Further
motivation behind the choice of the blazar--neutrino association is
explored through simple calculations in \citet{Capel:2021nf}.

We start by reviewing the results of \citetalias{IceCube:2018dnn} and
the motivation behind this work in Sect.~\ref{sec:IC18}, before
describing the modelling assumptions used in our simulations in
Sect.~\ref{sec:physics}. We then use these simulations to study the
probability of chance coincidences and the implications of a connected
neutrino and $\gamma$-ray flux in Sects.~\ref{sec:chance} and
\ref{sec:connection}, respectively. Finally, we discuss these results
in Sect.~\ref{sec:discussion} and summarise our conclusions in
Sect.~\ref{sec:conclusions}. We use $N$ to denote the number of
sources and $n$ to denote the number of neutrinos or photons.

\section{The TXS 0506+056--IC170922A association}
\label{sec:IC18}

The analysis reported in \citetalias{IceCube:2018dnn} uses a
likelihood-ratio test to quantify the significance of the
blazar--neutrino observation compared to the expectations under the
null hypothesis of no connection (see pages~S36--S41 of
\citetalias{IceCube:2018dnn}). Here, we summarise their procedure and
highlight the assumptions made and their importance for the
calculation of the significance. We make use of the following notation
when referring to both $\gamma$-ray and neutrino sources:
$\phi = \dd{n}/\dd{E}\dd{t}\dd{A}$ is the differential number flux,
$\Phi = \int_{E_\mathrm{min}}^{E_\mathrm{max}} \dd{E} \phi$ is the
flux in between $E_\mathrm{min}$ and $E_\mathrm{max}$ and
$F = \int_{E_\mathrm{min}}^{E_\mathrm{max}} \dd{E} E \phi$ is the
energy flux in the same energy interval.

A single neutrino event (IC170922A) is considered, along with
$N_\mathrm{src} = 2257$ catalogued, extragalactic Fermi-LAT
sources. The likelihood was not derived from first principles that
reflect our knowledge of the data-generating process, but it was
constructed in an heuristic way. It has the form of a mixture model
with signal and background components
\begin{equation}
  \mathcal{L} = n_s \mathcal{S} + (1 - n_s) \mathcal{B},
\end{equation}
where $n_s$ is the number of neutrino signal events which is either
$n_s = 0$ for the null hypothesis, or $n_s = 1$ for the signal
hypothesis. The signal contribution is taken to have the form
\begin{equation}
  \mathcal{S} = \sum_{i=1}^{N_\mathrm{src}} \mathcal{N}(\vec{x}_\nu | \vec{x}_i) w_i(t_\nu, \vec{x}_i),
  \label{eqn:signal_term}
\end{equation}
where
$\mathcal{N}(\vec{x}_\nu|\vec{x}_i) = \frac{1}{2 \pi \sigma^2} \exp
\Big(\frac{- |\vec{x}_i - \vec{x}_\nu|^2 }{2\sigma^2} \Big)$ is
multivariate normal distribution in two dimensions representing the
distribution of possible reconstructed event directions,
$\vec{x_\nu}$, given a source direction $\vec{x}_i$. The weight
$w_i(t_\nu, \vec{x}_i)$ is the contribution of source $i$ to the
neutrino signal at time $t_\nu$, which is simply the number of
expected events (see e.g. \citealt{Aartsen:2017km}), expressed as
\begin{equation}
  w_i(t_\nu, \vec{x}_i) = T_\mathrm{obs} \int_{E^\nu_\mathrm{min}}^{E^\nu_\mathrm{max}} \dd{E_\nu} \phi^\nu_i(E_\nu, t_\nu) A_\mathrm{eff}(E_\nu, \vec{x}_i),
\end{equation}
where $T_\mathrm{obs}$ is the observation time and
$A_\mathrm{eff}(E_\nu, \vec{x}_i)$ is the energy-dependent effective
area of IceCube in the direction of source $i$. If we assume that we
can define $\phi^\nu_i$ in terms of a normalisation
$\Phi^\nu_{0,i}(t_\nu)$ and fixed spectral shape $h_i(E_\nu),$ then we
have
\begin{equation}
  \begin{split}
    w_i(t_\nu, \vec{x}_i) & =  T_\mathrm{obs} \Phi^\nu_{0,i}(t_\nu) \int_{E^\nu_\mathrm{min}}^{E^\nu_\mathrm{max}} \dd{E_\nu} h_i(E_\nu) A_\mathrm{eff}(E_\nu, \vec{x}_i) \\ 
                          & = T_\mathrm{obs} [C w_{i,\mathrm{model}}(t_\nu)] w_\mathrm{acc}(\vec{x}_i).
  \end{split}\label{eqn:likelihood_weight_def}
\end{equation}
So, the $w_i$ term is split into a `model' weight that is proportional
to the expected neutrino flux and an `acceptance' weight that depends
on the convolution of the effective area and spectral shape. The
background contribution is defined by the directional distribution of
alert events that are due to background,
\begin{equation}
  \mathcal{B} = \frac{p(\vec{x}_\nu | \mathrm{BG})}{2\pi}.
\end{equation}
This distribution was constructed from a Monte Carlo simulation of a
large number of neutrino events, including both the atmospheric
contribution and the astrophysical contribution simulated according to
the diffuse flux results from \citet{Aartsen:2016xlq}. The test
statistic is defined as the log likelihood ratio of two fixed
hypotheses: $n_s=0$ or $n_s=1$ from the proposed Fermi-LAT
sources. This expression simplifies to
\begin{equation}
  \mathrm{TS} = 2 \log{\frac{\mathcal{S}}{\mathcal{B}}}. 
\end{equation}
The reported $p$-value was then calibrated by comparing the observed
TS for IC170922A to the distribution of the TS from many simulated
background events. It is clear from the above expressions that a
larger $w_{i,\mathrm{model}}(t_\nu)$ for sources that are
directionally consistent with IC170922A leads to a more significant
result.

\citetalias{IceCube:2018dnn} tested four different assumptions for
${w_{i,\mathrm{model}}(t_\nu)}$: 1)
${w_{i,\mathrm{model}} = F^\gamma_i(t_\nu)}$ between 1 and 100~GeV, 2)
${w_{i,\mathrm{model}} = \Phi^\gamma_i(t_\nu) / \langle \Phi^\gamma_i
  \rangle}$ between 1 and 100 GeV, 3)
${w_{i,\mathrm{model}} = F^\gamma_i(t_\nu)}$ between 100~GeV and
1~TeV, and 4) ${w_{i,\mathrm{model}} = 1}$. Model weights are
calculated over a 28-day window containing $t_\nu$ from Fermi-LAT or
MAGIC observations, depending on the energy range considered. The
highest post-trial significance of $3\sigma$ is found for cases 1 and
2, and case 3 following closely with $2.8\sigma$ for the most
conservative trial correction assumptions. For case 4, the post-trial
significance reduces to $\sim$~$1.4\sigma$. So, in order to report a
detection that is significant, it is necessary to assume a model in
which the neutrino and $\gamma$-ray fluxes are proportional (see also
e.g. \citealt{Franckowiak:2020kd}).

As the choice of model is key to the significance reported, it is
important to consider the assumptions implied by these models. For
cases 1 and 2, we have
\begin{equation}
  w_{i,\mathrm{model}} = F^\gamma_i(t_\nu) = \frac{F^\nu_i(t_\nu)}{Y_{\nu\gamma}} = \frac{\Phi^\nu_{0, i}}{Y_{\nu\gamma}} \int_{E^\nu_\mathrm{min}}^{E^\nu_\mathrm{max}} \dd{E_\nu} E_\nu~h_i(E_\nu) 
\end{equation}
where we have introduced $F^\nu = Y_{\nu\gamma} F^\gamma$ as the
proportionality between the neutrino and $\gamma$-ray energy fluxes
that motivates this model choice (see
Sect.~\ref{sec:blazar_nu_connection} for further details). In order to
satisfy the global proportionality to the expected number of detected
events as shown in Eq.~\ref{eqn:likelihood_weight_def}, it is
necessary to assume the following: $Y_{\nu\gamma}$ is the same for all
sources and
$\int_{E^\nu_\mathrm{min}}^{E^\nu_\mathrm{max}} \dd{E_\nu}
E_\nu~h_i(E_\nu)$ is the same for all sources. The latter statement
implies that all sources must have the same neutrino spectrum. Indeed,
in \citetalias{IceCube:2018dnn} a neutrino power law spectrum
$\propto E^{-2.13}$ is assumed in the calculation of $w_\mathrm{acc}$,
motivated by the results of the diffuse flux analysis reported
in~\citet{Aartsen:2016xlq}.

\begin{figure*}[ht]
  \centering
  \includegraphics[width=0.7\textwidth]{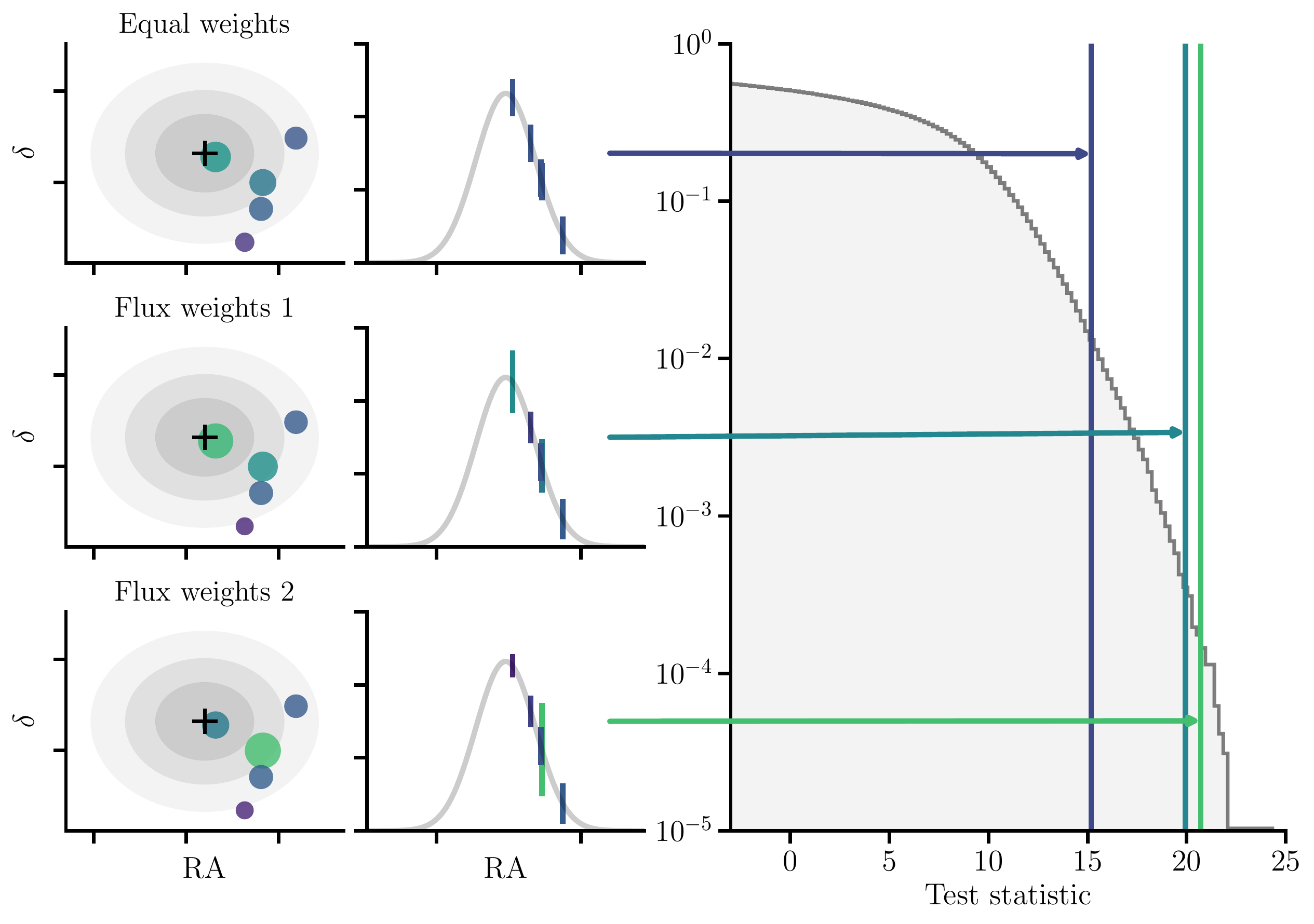}
  \caption{Overview of the likelihood-ratio test analysis used in
    \citetalias{IceCube:2018dnn}. The left column shows a
    reconstructed neutrino direction and its angular uncertainty on
    the sky with a cross and grey contours, respectively. Nearby
    source directions are shown by the blue circles, with darker and
    larger markers representing their contribution to the
    $\mathcal{S}$ term in Eq.~\ref{eqn:signal_term}. The centre column
    shows a 1D projection with the bars representing the flux weights,
    $w_{i,\mathrm{model}}$ for each source. The right panel shows the
    inverse cumulative test statistic distribution in grey, with the
    test statistics calculated in each case indicated by the
    arrows. The weights of nearby sources strongly influence the test
    statistic value and different scenarios can lead to significant
    results.}
  \label{fig:lhood_expl}
\end{figure*}

The method described above is summarised in Fig.~\ref{fig:lhood_expl},
which also illustrates how the values of $w_{i, model}$ can impact the
significance of a proposed association. Use of these weights is
reasonable when they are physically well-motivated. However, any rare
property of sources near to the neutrino localisation can result in a
significant result, even if the connection is
unphysical. Additionally, we see that multiple nearby sources can
contribute substantially to the final test statistic, whereas
physically we know that the neutrino event would have to come from a
single source. Generally, it is important to note that this
statistical approach does not address whether this neutrino is
actually produced by this source. Rather, it returns a significant
result if there are unusually many sources close to the neutrino
location with large weights. Thus, the method only addresses how
unlikely the observation is under the expectation of the background
model.

With this in mind, we model a population of flaring blazars based
purely on the available information in $\gamma$ rays to explore the
implications of this result. References to blazars and flares in the
text should be understood to mean $\gamma$-ray-bright blazars and
$\gamma$-ray flares, respectively.

\section{Simulation}
\label{sec:physics}

Our analysis is based on an empirical model of the blazar population,
the goal of which is to develop a simple framework that captures the
important features relevant to assess the implications of the proposed
neutrino association. A summary of all parameters introduced in this
section, as well as their reference values can be found in
Appendices~\ref{app:model_params} and \ref{app:alerts_model}. The code
used to implement the simulations can be found in the
\texttt{nu\_coincidence} repository
(\href{https://github.com/cescalara/nu_coincidence}{https://github.com/cescalara/nu\_coincidence}). For
the blazars we use the \texttt{popsynth} package
(\href{https://github.com/grburgess/popsynth}{https://github.com/grburgess/popsynth})
for population synthesis \citep{Burgess:2021kr}; for the neutrinos we
use the \texttt{icecube\_tools} package
(\href{https://github.com/cescalara/icecube_tools}{https://github.com/cescalara/icecube\_tools}). For
all simulations we assume a baseline joint observation period of
IceCube and Fermi of $T_\mathrm{obs} = 10$~years and a flat
$\Lambda$CDM cosmology with
$H_0 = 67.7$~km~$\mathrm{s}^{-1}$~$\mathrm{Mpc}^{-1}$,
$\Omega_m = 0.307$ and $\Omega_\Lambda = 1 - \Omega_m = 0.693$
\citep{Ade:2016md}.

\subsection{Blazar population}
\label{sec:blazar_pop}
  
There are two main categories of blazars, based on measurements of
their optical spectra: Flat-spectrum radio quasars (FSRQs) with strong
emission lines and BL~Lacertae-like objects (BL Lacs) with only weak
emission lines or featureless spectra \citep{Urry:1995ab}. The blazar
TXS~0506+056 was first classified as a BL~Lac, but upon closer
investigation it has been flagged as a likely `masquerading BL Lac',
so intrinsically an FSRQ, but with hidden emission lines
\citep{Padovani:2019wk}. In this way, we study both FSRQ and BL~Lac
populations here.

We consider the density of sources as a function of the rest frame
$\gamma$-ray luminosity in the 0.1--100 GeV range, $L_\gamma$, the
redshift, $z$, and the spectral index, $\Gamma$, such that
\begin{equation}
  \frac{ \dd{N} }{ \dd{L_\gamma} \dd{z} \dd{\Gamma}} = f_L(L_\gamma) f_\Gamma(\Gamma) \frac{\dd{N}}{\dd{V}} \frac{ \dd{V} }{ \dd{z} },
  \label{eqn:lum_func}
\end{equation}
where $f_L(L_\gamma)$ is the luminosity function, $f_\Gamma(\Gamma)$
is the distribution of spectral indices, $\dd{N}/\dd{V}$ is the source
density per comoving volume and $\dd{V}/\dd{z}$ is the comoving volume
element (see e.g. \citealt{Peacock:2010aa}). We assume that
$f_L(L_\gamma)$ does not evolve, similar to the `pure density
evolution' models presented in \citet{Ajello:2012kf,
  Ajello:2014lg}. This choice is motivated by the best-fit model found
in \citet{Marcotulli:2020mn}. Our results are also robust to the
possibility of an evolving luminosity function, as shown in
Appendix~\ref{app:lum_dep_evo}.

For both FSRQs and BL~Lacs, we model the luminosity function as a
broken power law
\begin{equation}
  f_L(L_\gamma) = \begin{cases} C_L L_\gamma^{-\alpha} & \mbox{for } L_\gamma \leq L_\mathrm{br},  \\  C_L L_\gamma^{-\beta} L_\mathrm{br}^{\alpha + \beta} & \mbox{for } L_\gamma > L_\mathrm{br}. \end{cases}
  \label{eqn:bpl}
\end{equation}
Here, $L_\mathrm{br}$ is the break luminosity and $C_L$ is defined
such that $f(L_\gamma)$ is normalised between
$L_{\gamma,\mathrm{min}}$ and $L_{\gamma,\mathrm{max}}$. The spectra
of these sources is assumed to be well-described by a power-law with a
single spectral index. For both populations, we use a normal
distribution model for the spectral index distribution
\begin{equation}
  f_\Gamma(\Gamma) = \frac{1}{\sigma_\Gamma \sqrt{2 \pi}} e^{-\frac{1}{2}(\frac{\Gamma - \mu_\Gamma}{\sigma_\Gamma})^2},
\end{equation}
where $\mu_\Gamma$ and $\sigma_\Gamma$ are the mean and standard
deviation, respectively. Based on the results of \citet{Ajello:2012kf,
  Ajello:2014lg}, we use different parametrisations for FSRQs and BL
Lacs. For FSRQs
\begin{equation}
  \frac{\dd{N}}{\dd{V}} \bigg\rvert_\mathrm{FSRQ} = \rho_0 \frac{1 + rz}{1+(z/p)^d},
\end{equation}
where $\rho_0$ is the local source density at $z=0$ and the other
parameters give the shape of the distribution \citep{Cole:2001uv}. For
BL~Lacs, we simply use
\begin{equation}
  \frac{\dd{N}}{\dd{V}} \bigg\rvert_\mathrm{BL~Lac} = \rho_0 (1 + z)^{-\delta}.
\end{equation}
Using this parametrisation, the total number of blazars is independent
of the luminosity and spectral index as these are normalised to
integrate to one. Therefore, the expected total number of objects in
the observable Universe is given by
\begin{equation}
  \bar{N}_\mathrm{tot} = \int_{z_\mathrm{min}}^{z_\mathrm{max}} \dd{z} \frac{\dd{N}}{\dd{V}} \frac{\dd{V}}{\dd{z}}.
\end{equation}

The ability of the Fermi-LAT to detect an individual blazar depends on
its luminosity, distance and spectral index. The Fermi-LAT instrument
is sensitive down to some minimum flux and sources with harder spectra
can be detected down to lower fluxes \citep{Abdo:2010hf}. In this
work, we implement this effect as a cut on the energy flux of an
object such that the probability of detection is modelled as
\begin{equation}
  \Omega(F_\gamma, \Gamma) = \begin{cases} 1 & \mbox{for } \Gamma - a\log_{10}(F_\gamma) > b, \\
                               0 & \mbox{elsewhere}, \end{cases}
                             \label{eqn:selection}
                           \end{equation}
                           where
                           $F_\gamma = L_\gamma / 4 \pi D_L^2(z)$, and
                           $a$ and $b$ describe the linear
                           boundary. This selection is made on the
                           \textit{observed} values of $F_\gamma$ and
                           $\Gamma$, including the observational
                           uncertainties. The \textit{true} values of
                           $F_\gamma$ and $\Gamma$ could fall below
                           this boundary.

                           We choose a reference set of parameters
                           reflecting the best-fit models of
                           \citet{Ajello:2012kf} and
                           \citet{Ajello:2014lg} that also gives a
                           $\gamma$-ray detected blazar population
                           that is consistent with the results
                           reported in the Fermi 4FGL
                           \citep{Abdollahi:2020nf} and 4LAC
                           \citep{Ajello:2020bw} catalogues (see
                           Appendix~\ref{app:model_params} for
                           details). We include unclassified blazars
                           in our modelling, assuming for the
                           reference case that 90\% of unclassified
                           blazars are actually BL Lacs, and the rest
                           are FSRQs. Later in Sect.~\ref{sec:chance},
                           we also consider alternative cases where we
                           completely disregard unclassified blazars,
                           or assume that their classification follows
                           the ratio of detected BL Lacs and FSRQs.

\begin{figure}[ht]
                             \centering
                             \includegraphics[width=\columnwidth]{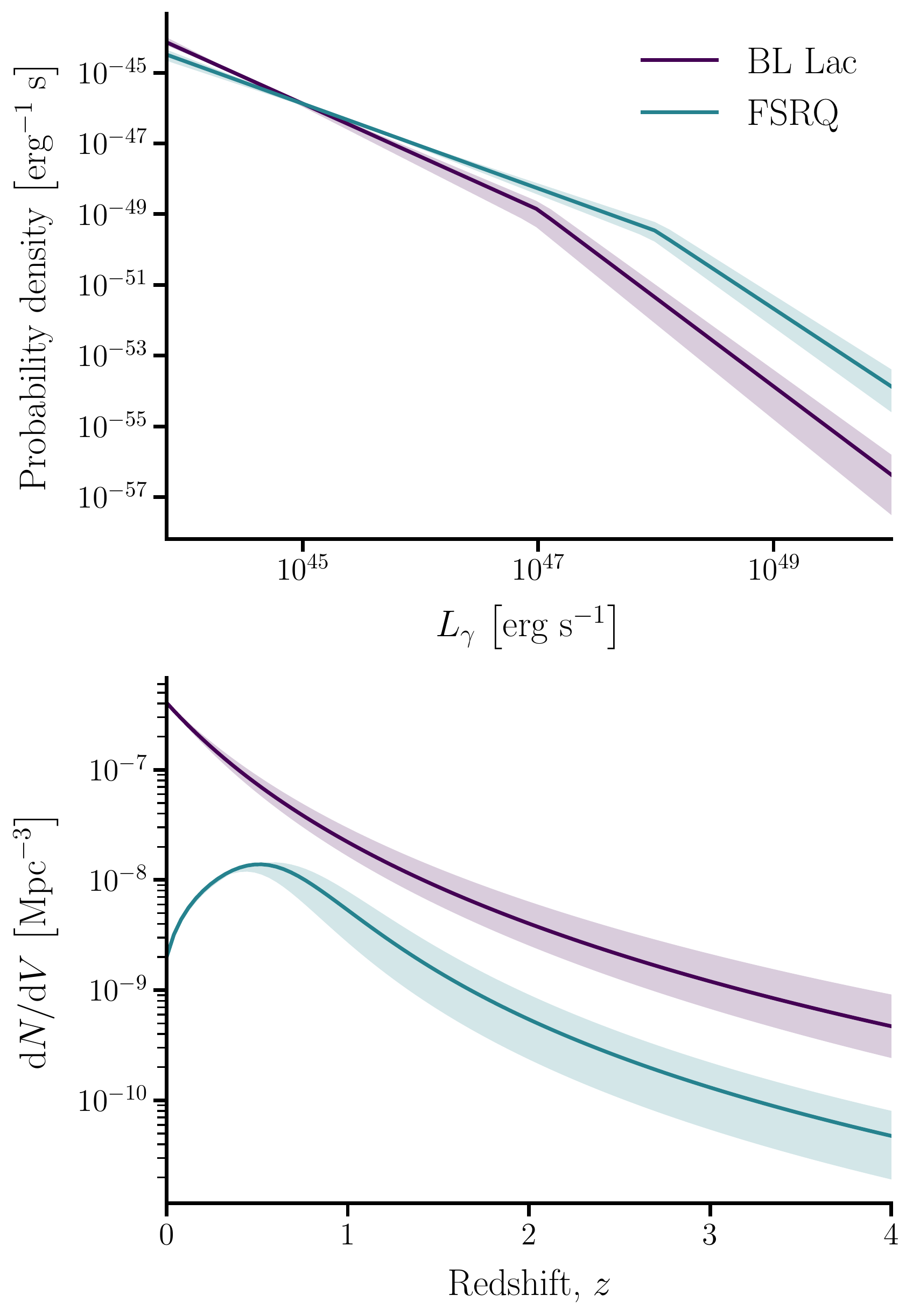}
                             \caption{Luminosity function (upper
                               panel) and the source density evolution
                               (lower panel) are shown for the range
                               of blazar population models tested. The
                               reference models are shown by solid
                               curves, and the shaded regions indicate
                               the bounds of the extreme models. The
                               density evolution can be compared to
                               Fig. 11 in \citealt{Ajello:2014lg},
                               although their results are for a
                               `luminosity-dependent density
                               evolution', and so should not be
                               expected to be exactly the same as our
                               simpler model.}
                             \label{fig:blazar_lum_evolution}
                           \end{figure}

\begin{figure}[h]
                             \centering
                             \includegraphics[width=\columnwidth]{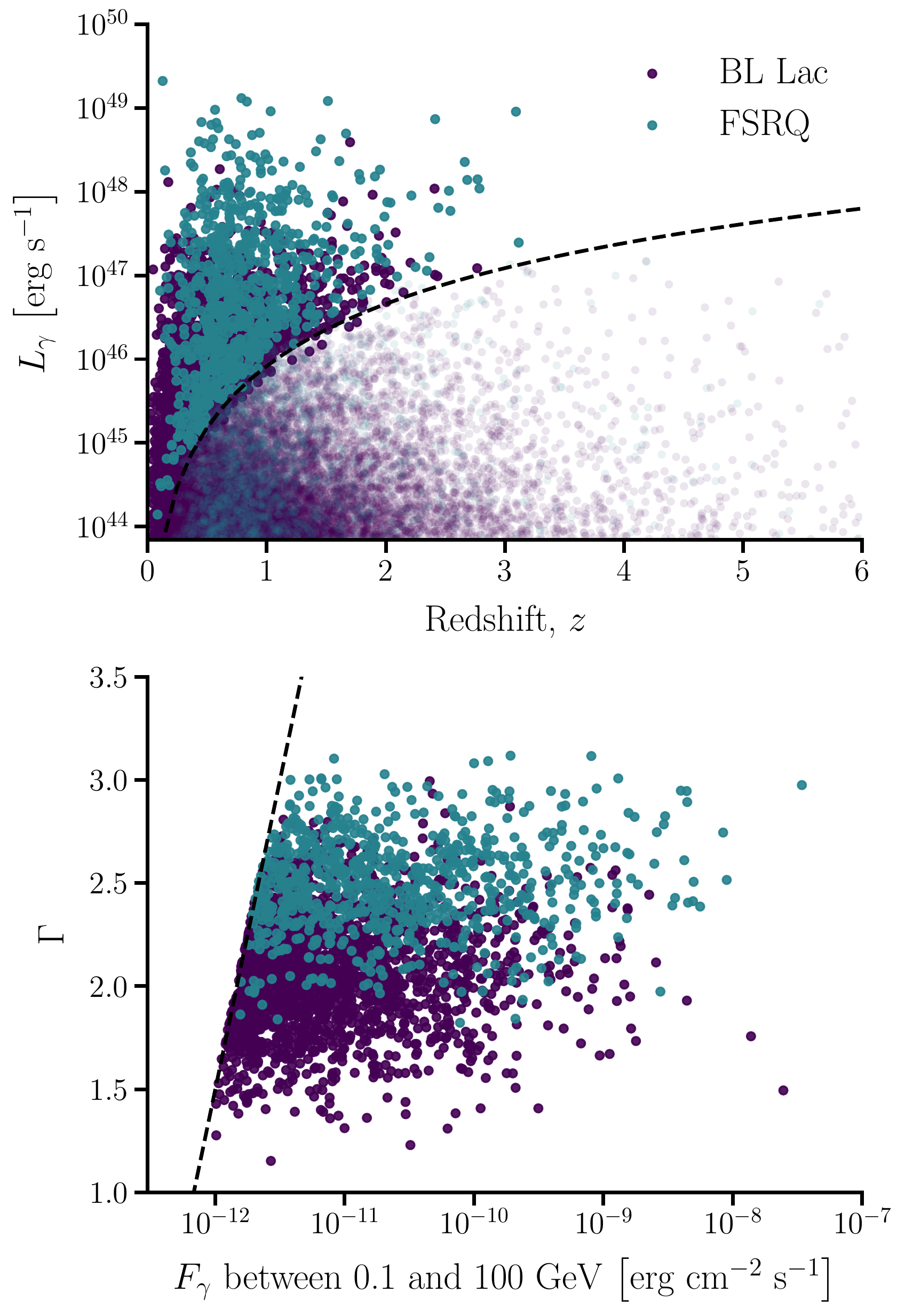}
                             \caption{Distributions of blazar
                               properties for an example simulated
                               blazar population assuming the
                               reference model. The upper panel shows
                               $L_\gamma$ and $z$ and the lower panel
                               shows $\Gamma$ and $F_\gamma$
                               (c.f. Fig.~4 in
                               \citealt{Ajello:2020bw}). In both
                               cases, the dashed lines show the
                               implemented selection. In the upper
                               panel, non-observed blazars are also
                               shown with transparent points. In this
                               particular simulation, there are a
                               total of 17\,551 sources, of which
                               2\,988 are detected (2\,206 BL~Lacs and
                               782 FSRQs).}
                             \label{fig:sim_blazar_dist}
                           \end{figure}

                           Motivated by the difficulty in estimating
                           the properties of the unknown blazar
                           populations, we also consider two extreme
                           cases of our BL Lac and FSRQ population
                           models that lead to lower and higher
                           numbers of detected sources, within
                           reasonable bounds ($\pm \sim 50\%$ of the
                           reference value). The luminosity function
                           and density evolution for these models are
                           shown in
                           Fig.~\ref{fig:blazar_lum_evolution}. The
                           distributions of blazar properties for an
                           example simulation under the reference
                           model are also shown in
                           Fig.~\ref{fig:sim_blazar_dist}.

                           \subsection{Blazar flares}
                           \label{sec:blazar_flare_model}

                           To model the flaring behaviour of blazars,
                           we use an empirical model that is based on
                           the results of the Fermi all-sky
                           variability analysis (FAVA,
                           \citealt{Abdollahi:2017fg}). FAVA is a
                           photometric analysis of the Fermi-LAT data
                           that searches for flux variations with a
                           time resolution of one week. The search is
                           carried out over two energy bands, from
                           0.1--0.8~GeV and 0.8--300~GeV. If
                           deviations of $> 6 \sigma$ in one energy
                           band or $> 4\sigma$ in both energy bands
                           with respect to the average are found, then
                           the flares are catalogued. The results are
                           updated in real-time, and the FAVA analysis
                           of TXS~0506+056 is one of the ways in which
                           it was identified as an active source.

                           We use the second FAVA catalogue with
                           7.5~years of data \citep{Abdollahi:2017fg}
                           as a basis for modelling the fraction of
                           variable blazars in a population, as well
                           as the rate, duration and amplitude of
                           significant $\gamma$-ray flares. The choice
                           of power-law distributions is motivated by
                           the observed values and the same
                           parametrisation is used for both BL~Lacs
                           and FSRQs. Blazar variability is seen to be
                           luminosity dependent, with more luminous
                           ($L_\gamma \gtrsim
                           10^{46}$~erg~$\mathrm{s}^{-1}$) objects
                           tending to be more variable, although this
                           is partly due to detection effects
                           \citep{Ackermann:2011js}. We do not model
                           this directly here, but instead use
                           different parameters to model the FSRQs as
                           more variable, resulting in a similar
                           effect as FSRQs tend to be brighter (see
                           Figs.~\ref{fig:blazar_lum_evolution} and
                           \ref{fig:sim_blazar_dist}).

                           All sources in a given population have a
                           probability to exhibit variable behaviour,
                           which we parametrise with the expected
                           fraction of variable sources,
                           $f_\mathrm{var}$. Sources that are not
                           variable have a flare rate of
                           $R_\mathrm{f} = 0$. The distribution of
                           flare rates for variable sources is given
                           by a bounded power law distribution
                           \begin{equation}
                             P(R_\mathrm{f} | f_\mathrm{var} > 0) = C_R R_\mathrm{f}^{-\eta_R},
                           \end{equation}
                           for
                           $R_\mathrm{f}^\mathrm{min} \leq
                           R_\mathrm{f} <
                           R_\mathrm{f}^\mathrm{max}$. Here $C_R$ is
                           defined such that the distribution is
                           normalised. The flare rate can then be used
                           to calculate the expected number of flares,
                           $\bar{N}_f$, in a given observation period,
                           $T_\mathrm{obs}$. Flares are assumed to
                           occur uniformly in time over the specified
                           $T_\mathrm{obs}$. Each flare has a
                           duration, $\tau$, that is also distributed
                           according to a bounded power law
                           \begin{equation}
                             P(\tau) = C_\tau \tau^{-\eta_\tau} \mbox{ for } \tau_\mathrm{min} \leq \tau < \tau_\mathrm{max},
                           \end{equation}
                           where $C_\tau$ is again defined to
                           normalise the distribution. The bounds,
                           $\tau_\mathrm{min}$ and
                           $\tau_\mathrm{max}$, are set in such a way
                           as to ensure a minimum duration of one week
                           and a maximum duration that does not result
                           in overlapping flares. Finally, each flare
                           also has an amplitude, $A_f$, defined as
                           the multiplicative factor by which the
                           luminosity is temporarily increased for the
                           flare duration. As we consider only large
                           flares here, such as would be counted as
                           significant by a FAVA analysis, we model
                           the distribution of amplitudes as a Pareto
                           distribution
                           \begin{equation}
                             P(A_f) = \frac{\eta_A {A_f^\mathrm{min}}^{\eta_A}}{A_f^{\eta_A + 1}},
                           \end{equation}
                           where $A_f^\mathrm{min}$ is the minimum
                           amplitude and $\eta_A > 0$ is the index.

                           This flare model is obviously a
                           simplification of the complexity present in
                           actual blazar lightcurves, with variability
                           seen over a wide range of timescales (see
                           \citealt{Boettcher:2019jf} for a recent
                           review). Ideally, we would start from a
                           physically motivated model that describes
                           such variable emission as a function of the
                           blazar properties, which could be convolved
                           with the blazar number density to calculate
                           the expected resulting contribution as a
                           function of time. However, in
                           \citetalias{IceCube:2018dnn}, blazar
                           lightcurves are binned into intervals of
                           one month when comparing with the time of
                           neutrino emission, so we choose a more
                           discrete flare model to reflect this. Given
                           the typical timescale of a few months for
                           flare durations, our `on/off' description
                           of blazar flares is a reasonable choice. We
                           note that by averaging over long periods,
                           short spikes in the emission that are
                           easier to detect due to lower instantaneous
                           background rates are not modelled and the
                           resulting flux is generally harder to
                           detect. This choice\footnote{We did
                             consider using a more sophisticated model
                             such as damped random walks or structure
                             functions \citep{Kozlowski:2016ns}, but
                             we would in any case attempt to tune
                             these models to reproduce the FAVA
                             results as a baseline.}  could have an
                           impact on the results presented in
                           Sect.~\ref{sec:blazar_nu_connection}.
  
                           \subsection{The blazar--neutrino
                             connection}
                           \label{sec:blazar_nu_connection}
  
                           To examine the case of a possible
                           blazar-neutrino connection, in this work we
                           consider two distinct approaches to
                           modelling the neutrino emission. Firstly,
                           we consider no connection between the
                           blazar population and the observed
                           high-energy neutrino flux. In this case,
                           the neutrino emission is modelled as an
                           isotropic, diffuse flux incident at the
                           Earth, at a level consistent with the
                           results of IceCube observations. This `null
                           model' allows us to evaluate the
                           probability for chance coincidences to be
                           present in this scenario, as detailed
                           further in Sect.~\ref{sec:chance}.

                           We also consider a direct connection
                           between the $\gamma$-ray and neutrino
                           emission, motivated by that assumed in
                           \citetalias{IceCube:2018dnn} to arrive at
                           the $3\sigma$ result, as discussed in
                           Sect.~\ref{sec:IC18}. In particular, we
                           assume that the energy flux in neutrinos is
                           some factor of that in gamma rays such that
                           \begin{equation}
                             F_\nu = Y_{\nu\gamma} F_\gamma,
                             \label{eqn:blazar_nu_connection}
                           \end{equation}
                           where $F_\nu$ is defined as the integral of
                           the flux between 10 TeV and 100 PeV, and
                           $F_\gamma$ is defined similarly between 1
                           and 100 GeV. Blazars in the population then
                           continuously produce neutrinos according to
                           Eq.~\ref{eqn:blazar_nu_connection}, and
                           their emission is amplified suitably during
                           flaring periods. This `connected model' is
                           used to examine the implications of an
                           assumed blazar--neutrino connection for the
                           blazar population in
                           Sect.~\ref{sec:connection}.

                           By adopting the relation stated in
                           Eq.~\ref{eqn:blazar_nu_connection}, we
                           assume that $Y_{\nu\gamma}$ is the same for
                           all blazars. This is not physically
                           motivated, as we would expect that the
                           ability of a blazar to produce neutrinos
                           would depend on its physical properties
                           (e.g. proton content and energies) that one
                           expects to vary from source to
                           source. Moreover, a connection between the
                           GeV photon flux and the TeV neutrino flux
                           might not be universally motivated by
                           physical assumptions. In single-zone hybrid
                           leptohadronic scenarios for TXS~0506+056
                           \citep{Keivani:2018ng, Gao:2019rn,
                             Cerruti:2019mj, Gasparyan:2022}, the GeV
                           photon flux primarily results from
                           synchrotron self-Compton of the
                           lower-energy synchrotron component. Thus,
                           the flux scales proportional to the product
                           of the number of electrons ($\nel$) and
                           low-energy photons ($\ngam$). On the other
                           hand, the flux of high-energy neutrinos
                           results from the interaction of accelerated
                           protons with the same low-energy
                           synchrotron photons; therefore, roughly
                           scaling as the product of the number of
                           protons ($\np$) and $\ngam$. This implies
                           that $\Yng \propto \frac{\np}{\nel}$. In
                           summary, the observed flux in GeV photons
                           and TeV neutrinos might not be due to the
                           same underlying processes, as possible
                           contributions to the GeV photon flux from
                           hadronic processes could be dominated by
                           those from purely leptonic
                           processes. Additionally, there is no reason
                           to believe that $\Yng$ remains constant
                           across a population of flares or even
                           within a single outburst. For
                           proton-synchrotron models, the GeV photon
                           flux scales proportional to $\np$ while the
                           TeV neutrino flux again scales proportional
                           to the product of $\np$ and the low-energy
                           $\ngam$. Therefore, $\Yng \propto \ngam$
                           which again should vary from source to
                           source and within a single outburst. Thus,
                           although $\Yng$ could be useful as an
                           observational tool, it is not a
                           well-motivated quantity for forward
                           simulation. Nevertheless, the weighting of
                           the likelihood used in
                           \citetalias{IceCube:2018dnn} implicitly
                           makes the assumption shown in
                           Eq.~\ref{eqn:blazar_nu_connection}, as
                           detailed in Sect.~\ref{sec:IC18}, and so we
                           investigate its implications here.

                           Another unphysical assumption in the
                           likelihood that we detailed in
                           Sect.~\ref{sec:IC18} is that the neutrinos
                           are always produced with a power-law
                           spectrum that has a fixed spectral
                           index. In most physical scenarios
                           appropriate for blazar flare emission, the
                           peak and spectral index of the neutrino
                           spectrum are determined by the distribution
                           of low-energy photons. Notably, this
                           results in a hard spectrum of $~ E^{-0.3}$
                           \citep{Padovani:2015ks, Gasparyan:2022},
                           very different to the assumption of
                           $E^{-2.13}$ used here (motivated by the
                           assumptions of
                           \citetalias{IceCube:2018dnn}). These
                           caveats should caution the reader that the
                           limits on $\Yng$ presented in
                           Sect.~\ref{sec:connection} should be
                           interpreted in the context of the original
                           likelihood assumptions and not that of a
                           full physical model.

                           \subsection{Neutrino observations}
                           \label{sec:nu_obs}

                           The IceCube neutrino observatory
                           reconstructs the energy and direction of
                           incoming neutrinos from secondary Cherenkov
                           radiation signals
                           \citep{Aartsen:2017tl}. An important
                           background to the study of astrophysical
                           neutrinos is the contribution of
                           atmospheric neutrinos, produced via cosmic
                           ray interactions in the Earth's
                           atmosphere. IceCube has set up a real-time
                           alerts system with the goal of using
                           multi-messenger observations to help
                           identify possible neutrino sources through
                           follow-up programmes
                           \citep{Aartsen:2017jf}. To optimise the
                           potential of this system, published alert
                           events are selected via cuts on the amount
                           of photons deposited in the detector, and
                           to favour track-like event topologies from
                           the charged-current interactions of muon
                           neutrinos. This results in the selection of
                           events with a reasonable likelihood of
                           being astrophysical and smaller angular
                           errors on the reconstructed
                           directions. These alerts come in two main
                           categories: the high-energy starting tracks
                           (HESE); and the extremely high-energy
                           tracks (EHE). This real-time alert system
                           led to the identification of IC170922A
                           reported in \citetalias{IceCube:2018dnn}.

                           In this work, we model the detection of
                           neutrino alerts to connect with the
                           analysis carried out in
                           \citetalias{IceCube:2018dnn}. We make use
                           of publicly available information on the
                           effective area of IceCube together with
                           sensible cuts on the reconstructed
                           deposited energy to build our detector
                           model, further details are given in
                           Appendix~\ref{app:alerts_model}. Since June
                           2019, the alerts system has been updated
                           from the HESE/EHE selections to the new and
                           improved astrotrack bronze and gold
                           selections reported in
                           \citet{Blaufuss:2019lu}. For the sake of
                           relevance to \citetalias{IceCube:2018dnn},
                           we only consider the HESE and EHE alerts in
                           our analysis.

                           For the diffuse neutrino emission simulated
                           in the null-model case, we model the
                           per-flavour astrophysical neutrino flux as
                           a power law with a flux normalisation of
                           $2 \times
                           10^{-18}$~$\mathrm{GeV}^{-1}$~$\mathrm{cm}^{-2}$~$\mathrm{s}^{-1}$~$\mathrm{sr}^{-1}$
                           and a spectral index of $2.6$. The
                           atmospheric component is modelled similarly
                           with a flux normalisation of
                           $5 \times
                           10^{-18}$~$\mathrm{GeV}^{-1}$~$\mathrm{cm}^{-2}$~$\mathrm{s}^{-1}$~$\mathrm{sr}^{-1}$
                           and a power-law index of 3.7 (reasonable
                           for the energies of $> 10$~TeV considered
                           here). These choices are based on those
                           assumed in \citet{Aartsen:2017jf} and
                           \citet{Kopper:2016fl}, to reproduce the
                           expected number of atmospheric and
                           astrophysical alerts each year. Further
                           information is given in
                           Appendix~\ref{app:alerts_model}.
 
                           \section{Chance coincidences}
                           \label{sec:chance}

                           We first consider the case where there is
                           no connection between blazars and
                           neutrinos, as described by the null model
                           introduced in
                           Sect.~\ref{sec:blazar_nu_connection}. The
                           goal of simulating this case is to
                           understand the level of chance coincidences
                           between blazars and neutrinos that can
                           occur, even if they are actually
                           unrelated. As described in
                           Sect.~\ref{sec:physics}, we generate random
                           sets of blazar populations and neutrino
                           observations. For each simulated survey, we
                           count the number of coincident
                           detections. Repeating this procedure
                           $\sim 10^4$ times, we report the fraction
                           of surveys which satisfy various
                           coincidence checks, which is effectively
                           the probability of such a search resulting
                           in a false positive.

                           To show how the number of chance
                           coincidences changes for different
                           searches, we consider different coincidence
                           criteria. We first define spatial
                           coincidence as a blazar position being
                           inside the 90\% confidence region of a
                           detected neutrino event, and temporal
                           coincidence as a neutrino arrival time
                           being during the flaring period of a
                           detected blazar. We then consider three
                           levels of coincidence: Spatial coincidence
                           of neutrinos with blazars (spatial),
                           spatial coincidence of neutrinos with
                           variable blazars (spatial + variable), and
                           spatial and temporal coincidence of
                           neutrinos with blazar flares (spatial +
                           variable + flare).

\begin{figure}[h]
  \centering
  \includegraphics[width=\columnwidth]{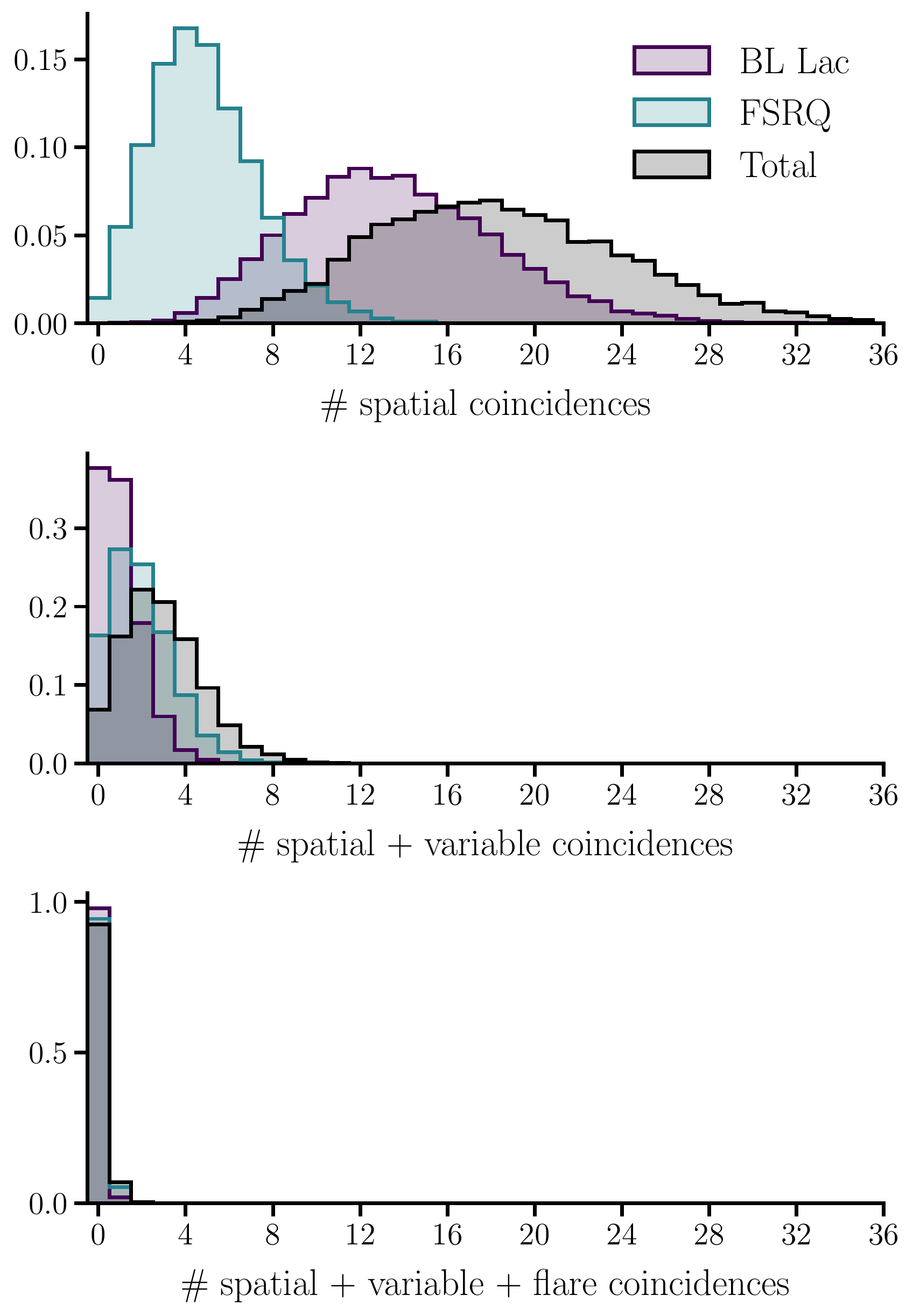}
  \caption{Distributions of the number of coincidences for simulations
    of the reference model given in
    Appendix~\ref{app:model_params}. Three difference coincidence
    levels are shown, as explained in the text. The BL Lac and FSRQ
    survey results are shown in purple and blue, respectively, with
    the total combined blazar survey shown in black.}
  \label{fig:coincidence_dist}
\end{figure}

                           For the reference blazar model parameters
                           given in Appendix~\ref{app:model_params},
                           the distributions of the number of
                           coincidences are shown in
                           Fig.~\ref{fig:coincidence_dist}. We see
                           that while BL~Lacs tend to have a larger
                           number of spatial coincidences as they are
                           more numerous, FSRQs have more variable
                           coincidences as they flare more often. In
                           general, we can expect observations of up
                           to around 32 spatial blazar coincidences
                           and 8 variable blazar coincidences to be
                           consistent with the null model. Flare
                           coincidences are rare, but not completely
                           unexpected. We find that 2.0\% and 5.6\% of
                           BL~Lac and FSRQ surveys of this size,
                           respectively, result in \textit{at least}
                           one flare coincidence. This gives a total
                           chance coincidence probability for flaring
                           blazars and neutrino alerts of 7.6\%. The
                           result is not affected much if we consider
                           the number of surveys leading to
                           \textit{exactly} one flare coincidence,
                           reducing to 7.4\% in this case.

                           Roughly half of these flare coincidences
                           will actually be due to the atmospheric
                           neutrino background, which contributes
                           4.5\% to the total 7.6\% for the reference
                           model case discussed above. For neutrino
                           events with energies in the range 200 TeV
                           -- 7.5 PeV, the 90\% confidence interval
                           found for IC170922A, (see Fig. S2 in
                           \citetalias{IceCube:2018dnn}), $\sim 37\%$
                           are due to the atmospheric background.

\begin{figure}[h]
  \centering
  \includegraphics[width=\columnwidth]{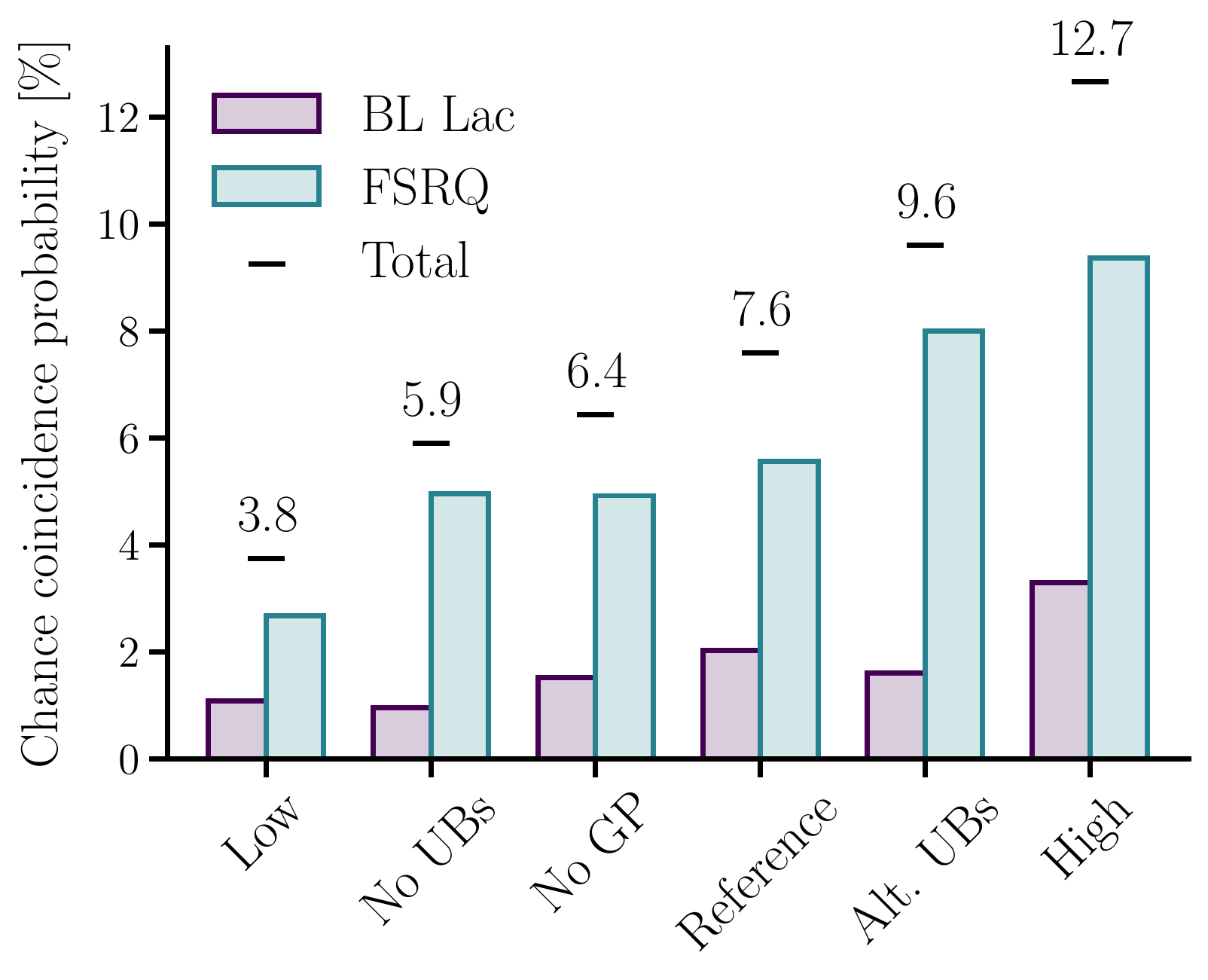}
  \caption{Chance coincidence probability of flare coincidences for
    the range of blazar population models shown in
    Fig.~\ref{fig:blazar_lum_evolution} and other variations described
    in the text. For each case, the separate contributions of BL~Lacs
    and FSRQs are shown in purple and blue, respectively.}
  \label{fig:coincidence_prob}
\end{figure}

Considering the higher and lower source density models that were
introduced in Sect.~\ref{sec:blazar_pop}, the chance coincidence
probabilities decrease and increase as expected relative to the number
of detected sources present in the simulation. For brevity, we focus
here on the interesting case of spatial + variable + flare
coincidences, most relevant to the results of
\citetalias{IceCube:2018dnn}. The chance coincidence probability for
these flare coincidences for the different population models is shown
in Fig.~\ref{fig:coincidence_prob}. Even considering rather large
changes in the population, we see that the total chance coincidence
probability is between 3.8\% and 12.7\%, with FSRQs as the dominant
contribution to this value.

To check the robustness of these results, we also consider further
variations to our reference blazar population model. We test excluding
$10^\circ$ either side of the Galactic plane in the selection of
detected blazars (No GP), motivated by the 4LAC catalogue
results. Additionally, we consider different treatment of the
unclassified blazars (UBs) reported in the Fermi surveys. For our
reference model, we assume that most (90\%) of all unclassified
blazars are BL Lacs. We also consider excluding unclassified blazars
completely (No UBs) or assuming that the ratio of FSRQs to BL Lacs is
the same as for classified blazars (Alt. UBs). The impact of these
assumptions is also shown in Fig.~\ref{fig:coincidence_prob}. Changing
the diffuse neutrino flux model also has a minor impact on the results
and this is detailed in Appendix~\ref{app:alerts_model}.
  
\begin{figure}[h]
  \centering
  \includegraphics[width=\columnwidth]{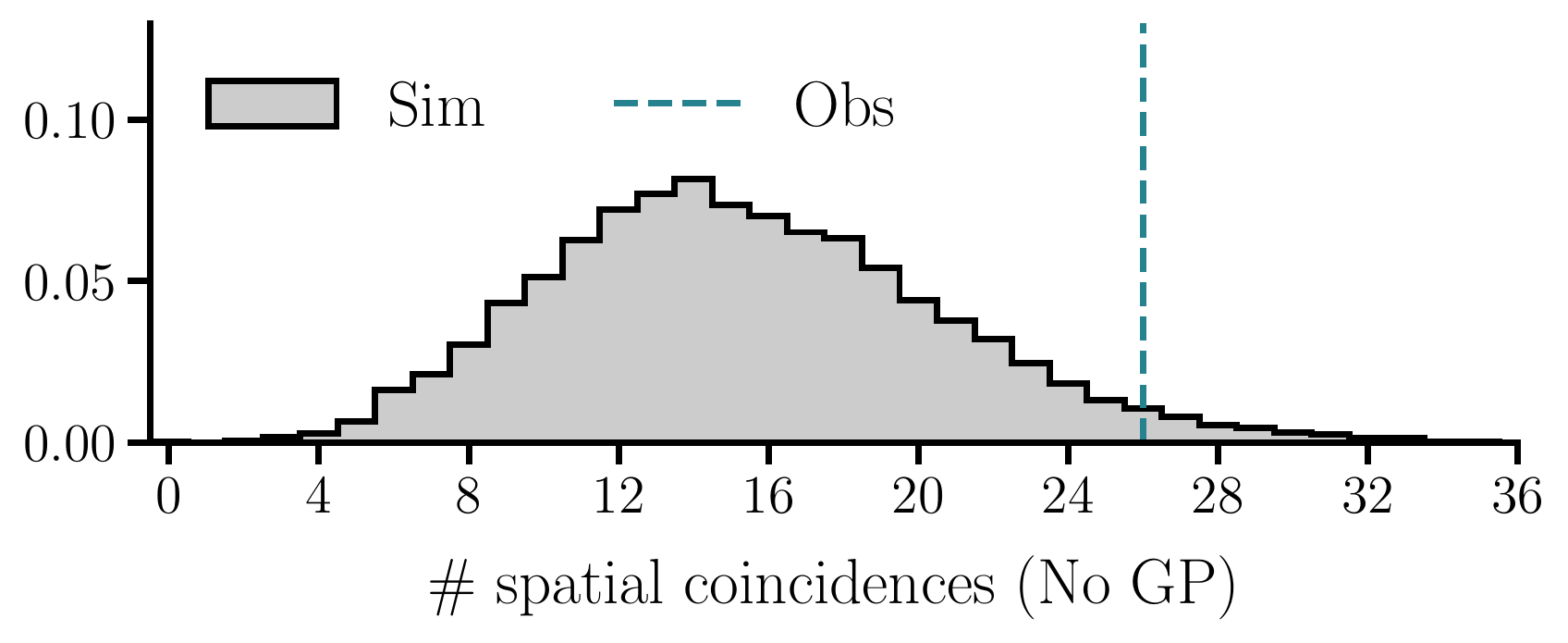}
  \caption{Distribution of the number of spatial coincidences found
    for the No GP case (histogram), to allow comparison with the
    observed number of 26 spatial coincidences reported in
    \citet{Giommi:2020md} (dashed line).}
  \label{fig:coincidence_dist_giommi}
\end{figure}

The expected value of spatial coincidences is typically less than that
seen in current searches, with 26 Fermi-detected blazar sources (15 BL
Lacs, six FSRQs, and five unclassified blazars) found within the 90\%
error region of IceCube alert events \citep{Giommi:2020md}\footnote{We
  consider sources that are present in the Fermi 4FGL catalogue and
  have a classification of either BL Lac, FSRQ or unclassified
  blazar. If we include all sources reported in \citet{Giommi:2020md},
  the total is 29.}. For a fair comparison with the assumptions used
in \citet{Giommi:2020md}, we must consider the number of spatial
coincidences for the No GP case introduced above, as shown in
Fig.~\ref{fig:coincidence_dist_giommi}. We find that the currently
observed number of spatial coincidences is more than the most probable
value, but consistent with the expectations of the null model. In our
null simulation we are most likely to see 14 coincidences, with a
probability of 0.08, and the probability of seeing 26 coincidences is
0.01\footnote{If we consider 29 observed events, this probability
  becomes 0.004}. This result is consistent with the conclusions of
\citet{Giommi:2020md}, as their most significant result is calculated
considering the intermediate and high energy peaked BL Lac sub
populations and neutrino event error regions that are enlarged by a
factor of 1.3 to account for possible underestimation, neither of
which are considered here.

\begin{figure}[h]
  \centering
  \includegraphics[width=\columnwidth]{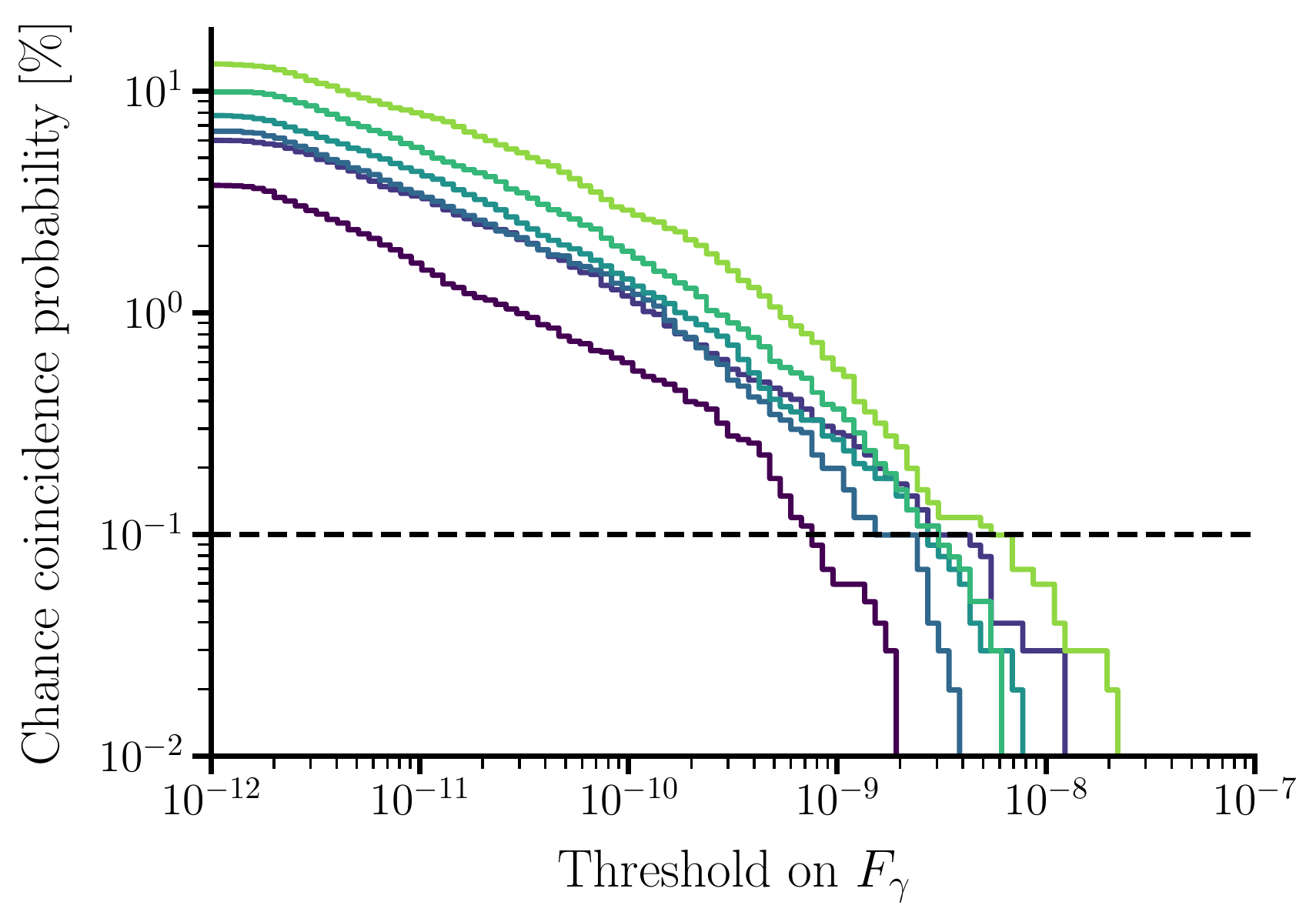}
  \includegraphics[width=\columnwidth]{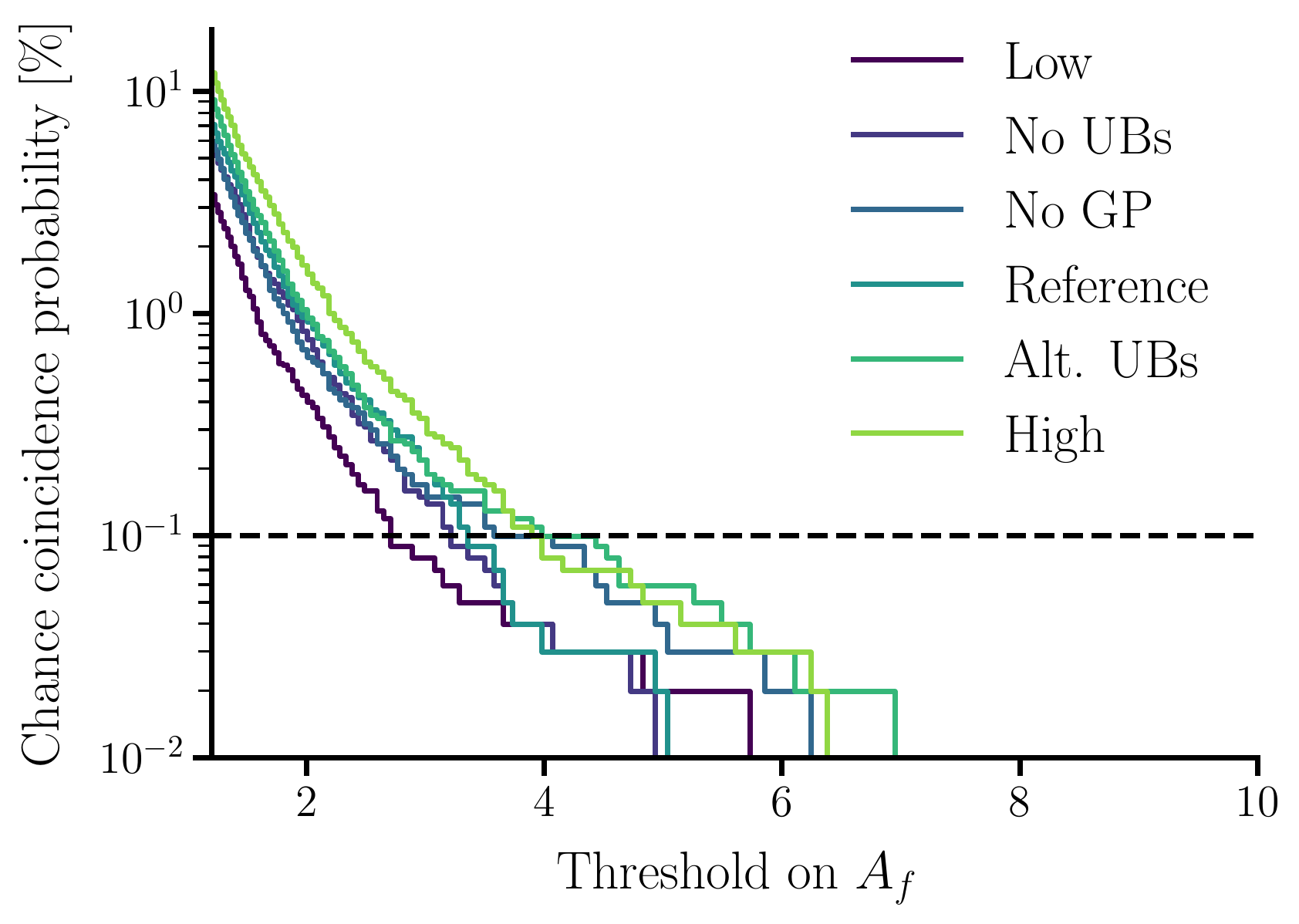}
  \caption{Chance coincidence probability as a function of the blazar
    flux (upper panel) and flare amplitude (lower panel)
    threshold. The different coloured lines show the different blazar
    population model assumptions, as in
    Fig.~\ref{fig:coincidence_prob}, and the dashed line at 0.1\%
    gives the $\sim 3\sigma$ level.}
  \label{fig:coincidence_prob_thresh}
\end{figure}

The chance coincidence probability that we find is small, but
non-negligible, and becomes increasingly problematic for longer or
deeper surveys. To connect with the $p$-value of 3$\sigma$ (i.e. a
chance coincidence probability of $\sim 0.1\%$) reported in
\citetalias{IceCube:2018dnn}, it is important to remember that their
calculation is tied to the model assumed in the likelihood for
$w_{i,\mathrm{model}}(t_\nu)$ as reviewed in
Sect.~\ref{sec:IC18}. When assuming no $\gamma$-ray--neutrino
connection, that is ${w_{i,\mathrm{model}}(t_\nu) = 1}$,
\citetalias{IceCube:2018dnn} also find a higher chance coincidence
probability. Assuming
${w_{i,\mathrm{model}}(t_\nu) = F_\gamma(t_\nu)}$ or
${w_{i,\mathrm{model}}(t_\nu) = \Phi_\gamma(t_\nu) / \langle
  \Phi_\gamma \rangle}$ is necessary to find a significant
association. To illustrate this effect, we set a threshold on
$F_\gamma$ and the flare amplitude required for a coincident
detection, and show how our chance coincidence probability changes as
a function of this threshold in
Fig.~\ref{fig:coincidence_prob_thresh}. Our goal here is not to
exactly reproduce the results of \citetalias{IceCube:2018dnn}, but to
consider the possibility of a blazar-neutrino connection from an
independent standpoint and compare the results in context.

Figure~\ref{fig:coincidence_prob_thresh} shows that we require a
blazar flux
$\gtrsim 10^{-9}$--$10^{-8}$~erg~$\mathrm{cm}^{-2}$~$\mathrm{s}^{-1}$
or a flare amplitude threshold $\gtrsim 3$--$5$ for a chance
coincidence probability of $< 0.1$\%. To give these values context,
for the reference model around 50 blazars in each survey have
$F_\gamma > 10^{-9}$~erg~$\mathrm{cm}^{-2}$~$\mathrm{s}^{-1}$ and the
flux of TXS~0506+056 during the 1 month period around the arrival of
IC170922A was
$3.3 \times 10^{-10}$~erg~$\mathrm{cm}^{-2}$~$\mathrm{s}^{-1}$
\citepalias{IceCube:2018dnn}. Similarly, about 50 flares in each
survey have $A > 3$ and the corresponding amplitude of the
TXS~0506+056 blazar flare over a 6 month period is $\sim 3.5$
\citepalias{IceCube:2018dnn}. The flare amplitude threshold is not
equivalent to $A_f^\mathrm{min}$ defined in
Sect.~\ref{sec:blazar_flare_model}, as it is a cut applied to the
simulated survey and not a parameter of the simulation itself.

To allow for a closer comparison with \citetalias{IceCube:2018dnn}, we
also implement the likelihood ratio test described therein and in
Sect.~\ref{sec:IC18}. For each simulated survey with one or more flare
coincidences, we find the test statistic distribution under the
assumption of the null hypothesis using $10^5$ simulated neutrino
alerts. We then compare the observed test statistic with this null
distribution to find the significance. When using a weighting based on
a linear relationship with the $\gamma$-ray flux, we confirm that we
find a chance coincidence probability of $\approx 0.13\%$ for events
with a test statistic threshold corresponding to $3\sigma$. This value
is as expected for a one-sided Gaussian definition of the $p$-value.

\section{Implications of a $\gamma$-ray--neutrino connection}
\label{sec:connection}

We now study the case of a connection between the integrated
$\gamma$-ray emission of blazars and the production of neutrinos,
introduced as the connected model in
Sect.~\ref{sec:blazar_nu_connection}. In \citetalias{IceCube:2018dnn},
$Y_{\nu\gamma}$ is estimated to be in the range of 0.5--1.7, assuming
a relevant neutrino energy range of 200 TeV to 7.5 PeV and depending
on the details of the neutrino emission timescale. However, taking
$Y_{\nu\gamma}$ and naively extrapolating to the whole blazar
population, we overshoot both the total number of neutrino alerts and
the number of alerts that share common sources \citep{Capel:2021nf}.

This overproduction can be accounted for if we adjust our estimate of
the neutrino flux necessary for the observation of a single alert
event. We expect that there are other sources which may contribute on
a similar level to that of TXS~0506+056. So, the expected contribution
from this individual source may be $\ll 1$, but the integrated
contribution from all sources in the population could be
$\mathcal{O}(1)$, as required for a detection
\citep{Strotjohann:2019nq}. Indeed, SED modelling of the
multiwavelength spectra of TXS~0506+056 indicate that it is
challenging to reach expected event numbers of $\mathcal{O}(1)$ for
this particular source, without overshooting the X-ray measurements or
invoking multi-zone models \citep{Keivani:2018ng, Cerruti:2019mj,
  Gao:2019rn, Xue:2019jk, Liu:2019mw}. Additionally, studies on the
modelling of similar blazar flares show expected neutrino event
numbers of $\ll 1$ for individual sources \citep{Oikonomou:2019me,
  Palladino:2019lc, Kreter:2020sk}. With this in mind, we allow for
the expected number of neutrino events from TXS~0506+056 to be $<1$
and explore the blazar--neutrino connection in this case.

\begin{figure}[h]
  \centering
  \includegraphics[width=\columnwidth]{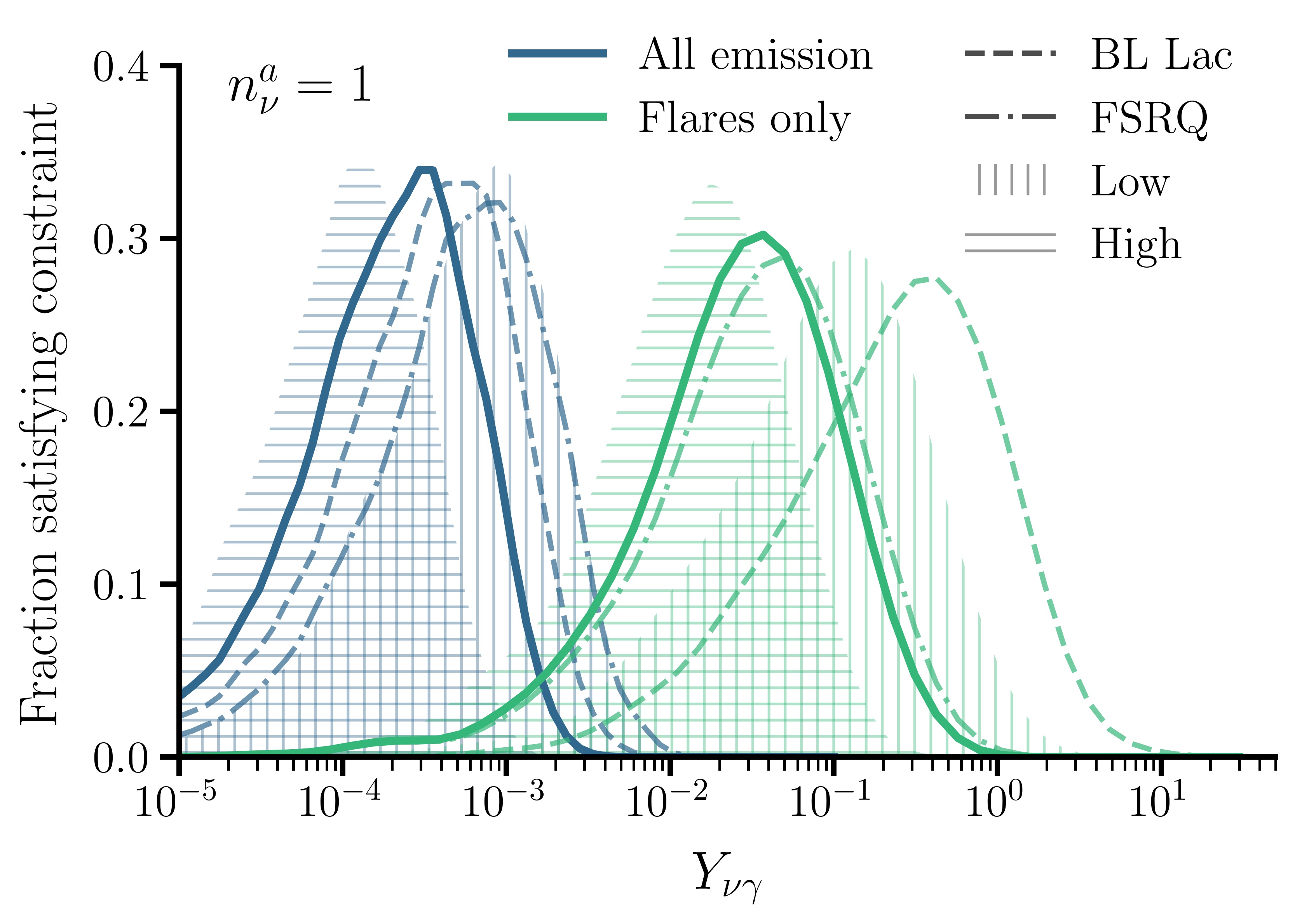}
  \caption{Fraction of simulated surveys satisfying $n_\nu^a = 1$ is
    shown for a range of $Y_{\nu\gamma}$ values. The blue lines show
    the results when considering both steady-state and flaring
    neutrino emission from blazars, and the green lines show the
    results from only flaring emission. For each case, the dashed and
    dash-dotted lines show the results for the BL~Lac and FSRQ
    populations, respectively. The hatched areas show the results for
    the higher and lower blazar population models introduced in
    Sect.~\ref{sec:blazar_pop}.}
  \label{fig:single_event_const}
\end{figure}

To estimate the constraints on $Y_{\nu\gamma}$ implied by the
observation of a single neutrino alert from the whole blazar
population, we ran $\sim 10^4$ simulations of our reference blazar
model for a range of different $Y_{\nu\gamma}$ values. We recorded the
fraction satisfying the constraint $n_\nu^a = 1$, where $n_\nu^a$ is
the detected number of neutrino alerts in IceCube after 10 years of
observations. The resulting constraints on $Y_{\nu\gamma}$ depend on
whether we considered steady-state neutrino emission from blazars or
purely flaring emission, as shown in
Fig.~\ref{fig:single_event_const}. We see that the constraints are
several orders of magnitude stronger when considering contributions
from all blazar emission, requiring $Y_{\nu\gamma} \lesssim
10^{-3}$. When considering only contributions from blazar flares to
the neutrino flux, $Y_{\nu\gamma} \sim 10^{-3}$--$10^{-1}$ is
consistent with $n_\nu^a = 1$. We can also see that this constraint is
dominated by the FSRQ population, as expected from the higher observed
variability in this case. Lowering the number of sources in the blazar
population relaxes these constraints, and vice versa.

\begin{figure}[h]
  \centering
  \includegraphics[width=\columnwidth]{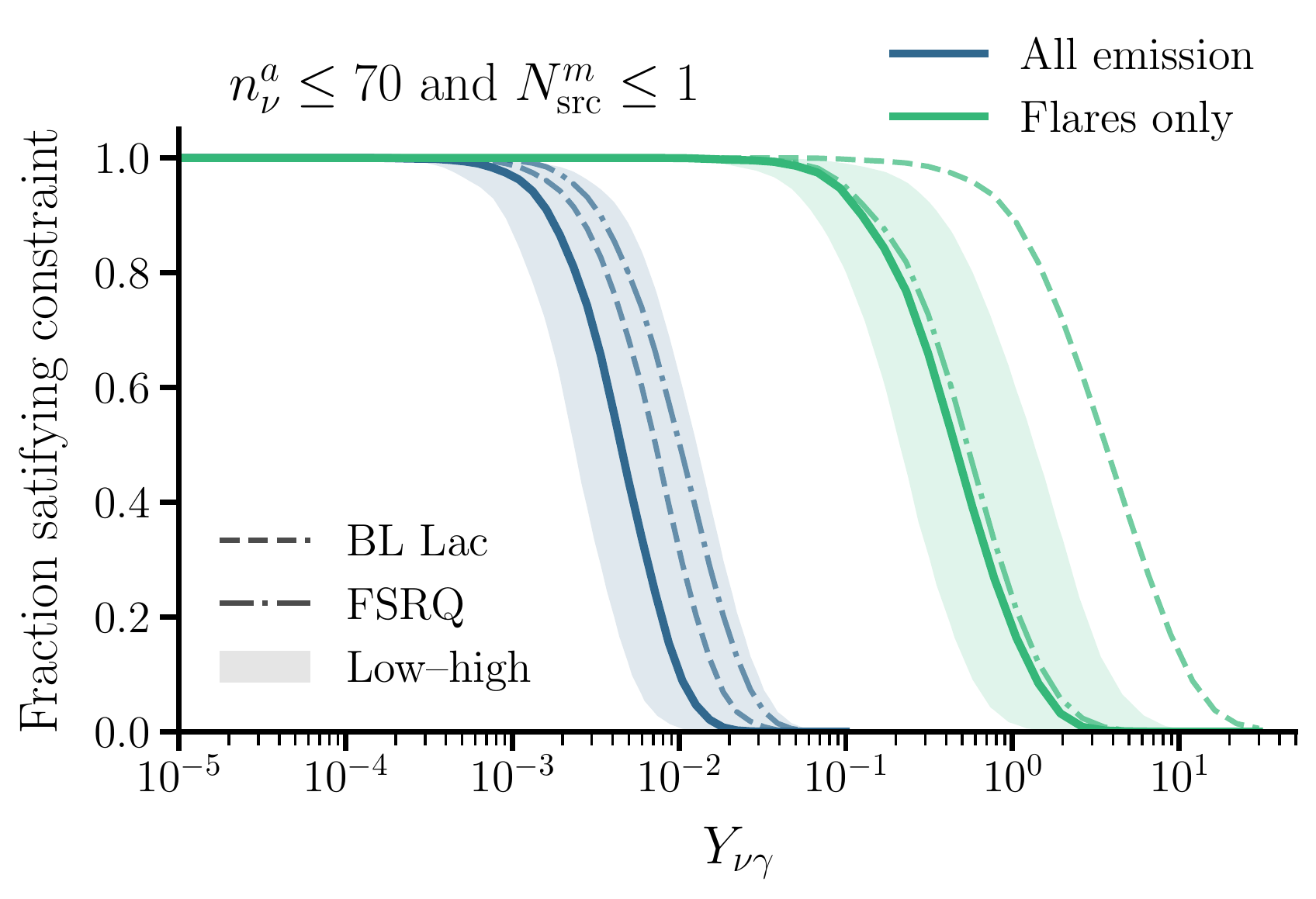}
  \caption{Fraction of simulated surveys satisfying $n_\nu^a \leq 70$
    and $N_\mathrm{src}^m \leq 1$ is shown for a range of
    $Y_{\nu\gamma}$ values. The different lines are as in
    Fig.~\ref{fig:single_event_const}, and the shaded regions give the
    bounds of the lower and higher population models.}
  \label{fig:total_event_const}
\end{figure}

We now also consider the more conservative constraints on
$Y_{\nu\gamma}$ from requiring $n_\nu^a \leq 70$, the total number of
observed neutrino alert events in 10~years\footnote{Based on the
  publicly available information, there are a total of 67 HESE/EHE
  alerts up until mid-2019, when the new system was introduced. Here
  we assume 70 events for the 2010--2020 period, but using 67 has
  negligible effect on Fig.~\ref{fig:total_event_const}.}, as reported
in \citetalias{IceCube:2018dnn} and the NASA GCN
archive\footnote{\href{https://gcn.gsfc.nasa.gov}{https://gcn.gsfc.nasa.gov}}. We
also know that we have yet to observe more than two neutrino alerts
that are consistent with a shared source direction\footnote{In the
  public HESE events, there are two events with larger angular
  uncertainties that are consistent with a shared source
  location.}. In this way, we also require $N_\mathrm{src}^m \leq 1$,
where $N_\mathrm{src}^m$ is the numbers of sources that produce more
than one detected neutrino event. The results are shown in
Fig.~\ref{fig:total_event_const}. Here, the upper limits are slightly
relaxed compared to those shown in Fig.~\ref{fig:single_event_const},
$Y_{\nu\gamma}$ is restricted to $\lesssim 10^{-2}$ when considering
all blazar emission and $\lesssim 1$ when only considering blazar
flares.

For all cases considered above, the contribution to neutrino alerts
from the undetected blazar population only really becomes relevant at
the level of 5--10\% when considering all blazar emission, both
steady-state and flaring. Practically all neutrino alerts from the
flaring population originate in blazars that would also be detected in
$\gamma$ rays. Whilst this is somewhat intuitive given the simplistic
connection between $\gamma$ rays and neutrinos assumed, it is
generally important to keep in mind the possible neutrino signal from
uncatalogued blazars when considering the implications of
observational constraints.

\section{Discussion}
\label{sec:discussion}

Using the simulation model described in Sect.~\ref{sec:physics}, we
demonstrated in Sect.~\ref{sec:chance} that the assumption of a
$\gamma$-ray--neutrino connection is necessary to find a low chance
coincidence probability for neutrino alerts and $\gamma$-ray flaring
blazars. We then explored the constraints on such a connection implied
by the blazar population in Sect.~\ref{sec:connection}. These
constraints are broadly in agreement with the predictions presented in
the framework of the blazar simplified view \citep{Padovani:2015ks}
and the independent constraint of $Y_{\nu\gamma} \leq 0.13$ from the
non-observation of events with deposited energy $\gtrsim 1$~PeV in 7
years of IceCube data \citep{Aartsen:2016ld, Aartsen:2017wk}. There
are also more detailed theoretical investigations of individual
blazars, for which the constraints on the hadronic fraction of
$\gamma$-ray emission are strong
\citep{Oikonomou:2019me,Palladino:2019lc,Kreter:2020sk}.

While relevant for the model presented in
\citetalias{IceCube:2018dnn}, the results in
Sect.~\ref{sec:connection} are dependent on the assumptions of our
connected $\gamma$-ray and neutrino model for blazars. The simplified
and unphysical form of this connection does not allow us to interpret
the constraints on $\Yng$ beyond the context of the IceCube likelihood
and its implied emission model. Considering a more physical
connection, it would be reasonable to assume that $\Yng$ varies both
across the blazar population and during flaring periods, weakening the
constraints shown here. Additionally, we would expect that the
neutrino emission follows a peaked spectrum that varies between
blazars. This expectation will also impact the results compared to the
assumption of a fixed $E^{-2.13}$ power-law neutrino emission for all
blazars used here, but the details of this effect will depend on the
location of the peak of the neutrino spectrum relative to the TeV --
PeV energy range observable by IceCube. Alleviating these issues would
require a more physically motivated likelihood and proper joint
fitting of neutrino and photon spectra with multi-messenger tools such
as \texttt{3ML} \citep{Vianello:2015} as well as exploring population
simulations with physically derived spectral models such as that
presented in \citet{Gasparyan:2022}.

It has been shown that the required target photon field for efficient
TeV--PeV neutrino production could lead to sources becoming opaque to
$\gamma$ rays in the 1--100~GeV range due to $\gamma\gamma$
interactions, and that cascades in the source environment would lead
to electromagnetic contributions down to KeV or MeV energies
\citep{Dermer:2007cs, Murase:2016ck, Reimer:2019ls, Das:2022le}. So,
we actually expect $\gamma$-ray emission in the Fermi-LAT energy range
to be suppressed during periods of increased neutrino emission, unless
the fundamental understanding of relativistic jets is wrong. In this
case, other wavelengths may be more indicative of a possible
blazar--neutrino connection, such as radio or X-rays (a review can be
found in \citealt{Giommi:2021kd}). In more recent work, a significant
correlation between neutrino hotspots in the Southern sky and blazars
in the Roma-BZCat catalogue has been reported in \citet{Buson:2022je},
while \citet{Liodakis:2022lk} conclude that we should not expect to be
able to identify significant correlations between radio flares in AGN
and neutrinos with the currently available data and methods.

Given the form of the likelihood described in Sect.~\ref{sec:IC18} and
the results presented in \citetalias{IceCube:2018dnn}, it is clear
that assuming no relation to the $\gamma$-ray flux, or an inversely
proportional relation would not lead to a significant association
between IC170922A and TXS~0506+056. While we plan to explore other
models in future, the main goal of this work is to study the choices
made in the \citetalias{IceCube:2018dnn} analysis that led to the
$3\sigma$ significance for the IC170922A--TXS~0506+056 and their
direct implications. As such, we focus on a linearly proportional
connection to $\gamma$ rays.

Several studies have investigated possible correlations of neutrinos
with known catalogues of $\gamma$-ray blazars using likelihood-based
stacking analyses, including weighting the source contribution by
$F_\gamma$ \citep{Aartsen:2017km, Huber:2019lg, Neronov:2017kf}. The
non-observation of significant correlations in the time-integrated
data is used to place upper limits on the contribution of these
sources to the observed neutrino flux of $\sim$10 -- 30\%, depending
on the model assumptions. Futhermore, the blazar contribution has also
been limited to $\sim 6$\% by an alternative approach considering
three proposed individual blazar associations
\citep{Bartos:2021mn}. Requiring that the total integrated flux from
the population is less than the total observed diffuse neutrino flux
leads to similar constraints on $Y_{\nu\gamma}$ to those shown in
Fig~\ref{fig:total_event_const}. Applying a full point source search
analysis to each simulated neutrino survey would likely yield even
stronger constraints on $Y_{\nu\gamma}$, due to the inclusion of
information lower-energy events, but this approach was not tested
here.

The IceCube Collaboration has also found a flare of lower energy
neutrinos at the $3.5\sigma$ significance level by looking into the
past data at the position of TXS~0506+056 \citep{IceCube:2018kf}. We
do not focus on this result in our work, as the analysis is
conditioned on the assumption that the IC170922A--TXS~0506+056
association is real in order to select this position in the sky. The
low-energy neutrino flare was found during a period where TXS~0506+056
was not in a $\gamma$-ray flaring state, despite the fact that a
$\gamma$-ray--neutrino connection is required for the original
high-energy neutrino association to be significant. To avoid the
logical inconsistency between the two analyses, we focus on studying
the IC170922A--TXS~0506+056 association and its implications in
isolation here.

\section{Conclusions}
\label{sec:conclusions}

We present a framework for studying the implications of co\-incident
multi-messenger detections using Monte Carlo simulations. Our approach
allows us to connect individual coincidences to the relevant source
populations and model assumptions. We applied this framework to the
case of a coincident high-energy neutrino and $\gamma$-ray flaring
blazar reported in \citetalias{IceCube:2018dnn}.

Assuming no connection between blazars and neutrinos, we find that
there is a 7.6\% chance of mistakenly finding high-energy neutrino
alerts that are coincident with $\gamma$-ray blazar flares in 10-year
surveys. This value ranges from 3.8\% to 12.7\% when considering
extreme cases of our blazar population model. To reduce this chance to
the level where an IC170922A--TXS~0506+056-like event has a $3\sigma$
significance (i.e. $\sim 0.1\%$), we must also consider the blazar
flux or flare amplitude at the time of the neutrino event, similar to
the likelihood weighting used in \citetalias{IceCube:2018dnn} and
reviewed in Sect.~\ref{sec:IC18}. We also show that we expect to see
as many as $\sim 32$ directional blazar--neutrino chance coincidences
in a 10-year survey and we verified that this value is consistent with
current observations.

Considering that a linearly proportional $\gamma$-ray--neutrino flux
connection is required for the IC170922A--TXS~0506+056 to be
statistically significant, we then explored the implications of this
assumption for the blazar population as a whole. We find that either
the $\gamma$-ray--neutrino connection is restricted to
$Y_{\nu\gamma} \lesssim 10^{-2}$, or that only a small fraction of
blazars contribute to the neutrino flux. However, in
\citet{Capel:2020cj}, we demonstrate that rare and luminous blazars
capable of fulfilling these scenarios are also strongly constrained by
the non-observation of point sources, further complicating the
physical picture (consistent with
\citealt{Yuan:2020jw}). Alternatively, if we only consider
contributions from blazar flares, we expect
$Y_{\nu\gamma} \sim 10^{-3}$--$10^{-1}$ to be consistent with one
neutrino--blazar flare coincidence in 10 years of observations. The
constraints on $\Yng$ should be interpreted in the context of the
likelihood model assumed in \citetalias{IceCube:2018dnn} and described
in Sect.~\ref{sec:IC18}, rather than that of a more detailed physical
model.

The simplified nature of the models for neutrino association that are
built into the likelihoods used in \citetalias{IceCube:2018dnn} and
other IceCube point source searches mean that finding an object with a
rare property that lies within a neutrino error region can lead to a
significant result. When interpreting such a result, it is important
to understand if the rare property is physically well-motivated. Here,
we explore the implications of the model assumed in
\citetalias{IceCube:2018dnn} for a population of blazars and find that
the expected neutrino signals are overproduced with respect to
observations, unless small values of $\Yng$ are considered, or only
blazar flares contribute to neutrino emission. Both of these results
motivate changes to the original likelihood used: If $\Yng$ is very
small, the $\gamma$-ray flux is likely not the most useful weight to
use. If only blazar flares contribute, non-flaring sources should not
be considered. In future, the use of more physical, interpretable
models in association analyses can help to resolve this confusion.

Our approach can be used to study similar multi-messenger detections,
inform the logical consistency of likelihood models used in such
searches and aid in the design of targeted follow-up programmes. By
defining a detailed generative model that connects source populations
to neutrino observations, we also lay the foundation for performing
inference of population parameters from observations in this setting.

\section*{Software}
\texttt{NumPy} \citep{harris2020array}, \texttt{SciPy}
\citep{2020SciPy-NMeth}, \texttt{Astropy} \citep{astropy:2013,
  astropy:2018}, \texttt{Matplotlib} \citep{Hunter:2007},
\texttt{h5py} \citep{collette_python_hdf5_2014}, \texttt{Joblib}
\citep{joblib:2021ng}, \texttt{popsynth} \citep{Burgess:2021kr}

\begin{acknowledgements}
  
  We thank the Max Planck Computing and Data Facility for the use of
  the Raven HPC system. F. Capel acknowledges financial support from
  the Excellence Cluster ORIGINS, which is funded by the Deutsche
  Forschungsgemeinschaft (DFG, German Research Foundation) under
  Germany’s Excellence Strategy - EXC-2094-390783311. J. M. Burgess
  acknowledges financial support by the Deutsche
  Forschungsgemeinschaft (SFB 1258). We thank Damien B\'{e}gu\'{e} for
  useful conversations on the physics of neutrino production.
  
\end{acknowledgements}

\bibliographystyle{aa} \bibliography{blazar_nu}

\appendix
 
\section{Blazar model parameters}
\label{app:model_params}

In Table~\ref{tab:model_params_full} we list in full the values for
the reference BL~Lac and FSRQ populations used in this work, based on
the parametrisation introduced in Sect.~\ref{sec:physics}. These
parameters were chosen by starting with an equivalent density
evolution to that found in \citet{Ajello:2012kf, Ajello:2014lg}, and
comparing the results of simulations to the 4FGL
\citep{Abdollahi:2020nf}, 4LAC \citep{Ajello:2020bw}, and FAVA
\citep{Abdollahi:2017fg} catalogues. The parameters were then tuned to
find a reasonable match for the number of detected sources, the number
of flaring sources, and the total number of flares, along with the
distributions of all observed properties.

We also modelled the detection uncertainties where relevant. For
$F_\gamma$, these uncertainties follow a log normal distribution
centred on the latent value, and with a standard deviation of 0.1, and
similarly for $\Gamma$ we used a normal distribution with standard
deviation of 0.1. These choices are based on the values reported in
\cite{Abdollahi:2020nf}.

To study the impact of our assumptions, we considered variations to
this blazar reference model, as detailed in
Sects.~\ref{sec:blazar_pop} and \ref{sec:chance}.  While we do not
reproduce an exhaustive list of parameter values here, all choices are
supplied as YAML configuration files in the \texttt{nu\_coincidence}
repository that can easily be loaded into the main code for
reproducibility of the results.

\begin{table}[h]
  \centering
  \caption{Reference blazar model parameters.}
  \begin{tabular}{llll}
    \toprule
    & & \textbf{BL Lac} & \textbf{FSRQ} \\
    \midrule
    \multirow{3}{*}{\textbf{Luminosity}} & $\alpha$ & 1.5 & 1.2 \\
    & $\beta$ & 2.5 & 2.2 \\
    & $L_\mathrm{br}$ &  $10^{47}$ & $10^{48}$ \\
    \midrule
    \multirow{2}{*}{\textbf{Spectrum}} & $\mu_\Gamma$ & 2.1 & 2.5 \\
    & $\sigma_\Gamma$ & 0.25 & 0.20 \\
    \midrule
    \multirow{5}{*}{\textbf{Evolution}} & $\rho_0$ & 5100 & 25 \\
    & $r$ & -- & 15.0 \\
    & $d$ & -- & 4.5 \\
    & $p$ & -- & 0.7 \\
    & $\delta$ & 4.2 & -- \\
    \midrule
    \multirow{3}{*}[-6pt]{\textbf{Flares}} & $f_\mathrm{var}$ & 0.08 & 0.40 \\
    & $\eta_R$ & 2.2 & 2.0 \\
    & $\eta_\tau$ & 2.2 & 2.0 \\
    & $\eta_A$ & 4.5 & 4.0 \\
    \midrule
    \multirow{2}{*}{\textbf{Selection}} & $a$ & \multicolumn{2}{c}{3.0} \\
    & $b$ & \multicolumn{2}{c}{37.5} \\
    \bottomrule
  \end{tabular}
  \tablefoot{All quantities are dimensionless other than
    $L_\mathrm{br}$ with units of erg~$\mathrm{s}^{-1}$ and $\rho_0$
    with units of $\mathrm{Gpc}^{-3}$~sr (assuming $4\pi$ sky
    coverage). The minimum flare amplitude $A_f^\mathrm{min}$, is
    1.2. We also use
    $L_\mathrm{min} = 7 \times 10^{43}$~erg~$\mathrm{s}^{-1}$,
    $L_\mathrm{max} = 10^{50}$~erg~$\mathrm{s}^{-1}$,
    $z_\mathrm{min} = 0$, $z_\mathrm{max} = 6$,
    $R_\mathrm{f}^\mathrm{min} = 0.1$~$\mathrm{yr}^{-1}$, and
    $R_\mathrm{f}^\mathrm{max} = 10$~$\mathrm{yr}^{-1}$ throughout for
    both BL~Lacs and FSRQs.}
  \label{tab:model_params_full}
\end{table}

\section{IceCube real-time alerts model}
\label{app:alerts_model}

\begin{figure}[h]
  \centering
  \includegraphics[width=\columnwidth]{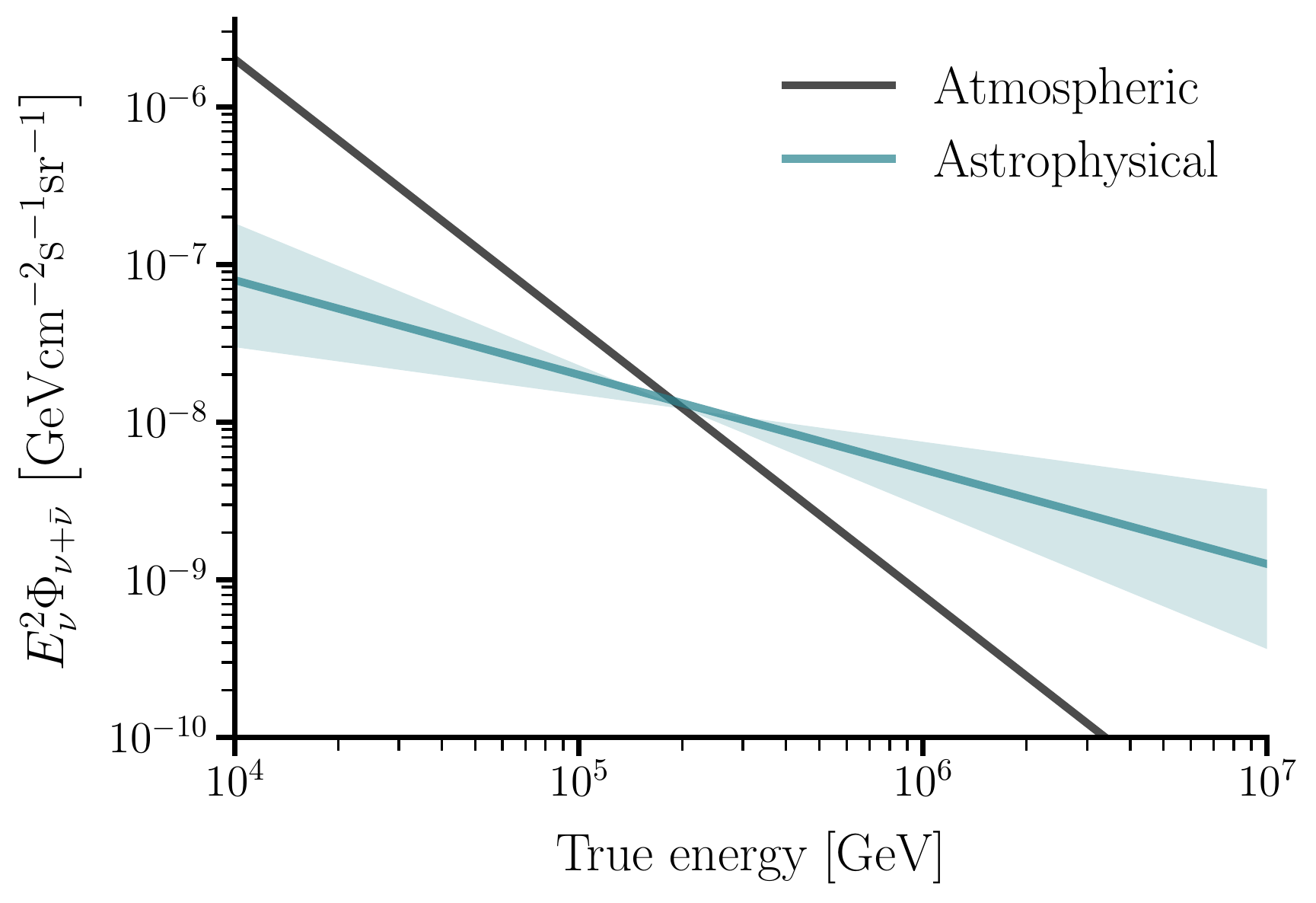}
  \caption{Per-flavour diffuse neutrino flux used is shown for both
    the atmospheric and astrophysical components. For the
    astrophysical case, the two extreme cases also considered bound
    the shaded region.}
  \label{fig:diffuse_nu_model}
\end{figure}

We modelled the diffuse neutrino flux as described in
Sect.~\ref{sec:nu_obs} and shown in
Fig.~\ref{fig:diffuse_nu_model}. The ratio of the atmospheric and
astrophysical factors is the driving factor behind the classification
of alert events, and we see that higher energy events are more likely
to be of astrophysical than atmospheric origin. The characterisation
of the astrophysical flux is still uncertain (see e.g. Fig.~5 in
\citealt{Abbasi:2022kd}). To investigate the effect of this
uncertainty, we also considered two extreme cases, as shown by the
shaded region in Fig.~\ref{fig:diffuse_nu_model}. These extreme cases
describe harder and softer flux models, the details of which are given
in Table~\ref{tab:diffuse_nu_params}. The uncertainties on the
atmospheric component are considered to be negligible in comparison to
those of the astrophysical component. In
Table~\ref{tab:diffuse_nu_params}, we also give the chance coincidence
probabilities for spatial + variable + flare coincidences, as studied
in Sect.~\ref{sec:chance}.

\begin{table}[h]
  \centering
  \caption{Summary of the per-flavour diffuse neutrino flux models
    considered.}
  \begin{tabular}{llll}
    \toprule
    \multirow{2}{0.7cm}{\textbf{Model}} & \multirow{2}{2.0cm}{\textbf{Flux normalisation}} & \multirow{2}{1.3cm}{\textbf{Spectral index}} & \multirow{2}{2.3cm}{\textbf{Chance coin. prob. (Astro. $\nu$)}} \\
    \\
    \midrule
    \textbf{Hard} & $1.5 \times 10^{-18}$ & 2.3 & 8.5\% (3.7\%) \\
    \textbf{Reference} & $2.0 \times 10^{-18}$ & 2.6 & 7.6\% (3.1\%) \\
    \textbf{Soft} & $2.3 \times 10^{-18}$ & 2.9 & 7.6\% (2.9\%) \\
    \bottomrule
  \end{tabular}
  \tablefoot{The flux normalisation is defined at 100~TeV and given in
    units of
    $\mathrm{GeV}^{-1}$~$\mathrm{cm}^{-2}$~$\mathrm{s}^{-1}$~$\mathrm{sr}^{-1}$. The
    chance coincidence probabilities are given for all neutrino alerts
    and in parentheses when only considering alerts of astrophysical
    origin. These results can be compared the reference case in
    Fig.~\ref{fig:coincidence_prob}.}
  \label{tab:diffuse_nu_params}
\end{table}

\begin{figure}[h]
  \centering
  \includegraphics[width=\columnwidth]{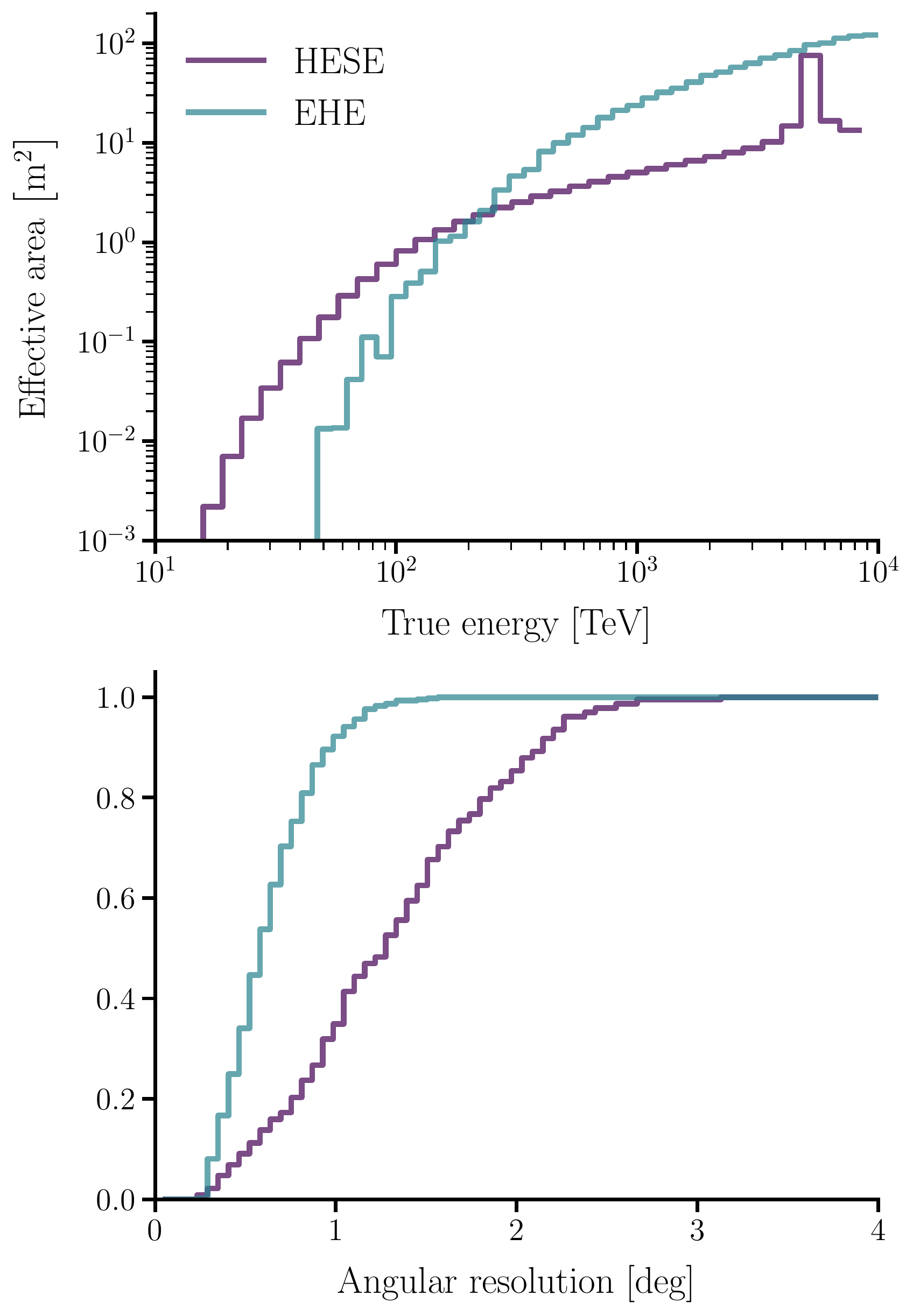}
  \caption{Details of the neutrino alert simulations. Left panel:
    Whole-sky effective area for the HESE and EHE track alerts
    generated via simulations using the detector mode described in
    Sect.~\ref{sec:nu_obs}. Right panel: Cumulative distribution for
    the angular resolution of simulated alert events (c.f. Figs.~7 and
    9 in \citealt{Aartsen:2017jf}).}
  \label{fig:alert_detector_model}
\end{figure}

To model the HESE alerts, we used the effective areas from the public
dataset associated with \citet{Aartsen:2013ps}, summed over all
flavours. Similarly, for the EHE alerts, we used the effective areas
from \citet{IceCube:2018nf} for muon neutrino track events. In both
cases, we reduced the effective areas slightly to account for the more
stringent event cuts described in \citet{Aartsen:2017jf}, and we
modelled the energy resolution using information from the data release
associated with \citet{Aartsen:2015fk}. For the EHE case, we placed a
cut on the reconstructed energies of $E_\mathrm{reco} > 250$~TeV to
match the requirements of the alerts stream. For the angular
resolution, we again made use of the \citet{IceCube:2018nf} dataset,
but with small adjustments made such that the resulting 90\%
confidence regions match what is expected based on
\citet{Aartsen:2017jf} and
\citet{IceCube:2018nv}. Fig.~\ref{fig:alert_detector_model} shows the
HESE and EHE effective areas. To find the effective area, we ran
simulations of $10^6$ HESE and EHE events from a known power-law
spectrum using the setup described in
Sect.~\ref{sec:nu_obs}. Fig.~\ref{fig:alert_detector_model} shows the
cumulative distribution of the angular resolution for HESE and EHE
events.

\section{Luminosity-dependent density evolution}
\label{app:lum_dep_evo}

\citet{Ajello:2012kf, Ajello:2014lg} find that a luminosity-dependent
density evolution (LDDE) best describes the BL Lac and FSRQ luminosity
functions. As described in Sect~\ref{sec:blazar_pop}, we use a simple
pure density evolution (PDE) model, motivated by the more recent
results of \cite{Marcotulli:2020mn} for all blazars. As the evolution
is an important factor in exploring the connection between blazar
populations and neutrinos (see e.g. \citealt{Neronov2020:kj,
  Yuan:2020jw, Capel:2020cj}), we describe the differences in these
models and their impact on our conclusions here.

The dependence of the blazar luminosity function on $L_\gamma$ and $z$
can be summarised as $f(L_\gamma, z) = \dd{N} / \dd{L_\gamma}
\dd{V}$. For the PDE case, the shape of the luminosity distribution is
independent of the cosmological evolution of the source density, such
that
\begin{equation}
  f_\mathrm{PDE}(L_\gamma, z) = f_L(L_\gamma) \frac{\dd{N}}{\dd{V}}(z),
\end{equation}
as stated in Eq.~\ref{eqn:lum_func} and shown in
Fig.~\ref{fig:blazar_lum_evolution}. For the LDDE case, $f$ does not
factorise and the shape of the cosmological density evolution is a
function of $L_\gamma$
\begin{equation}
  f_\mathrm{LDDE}(L_\gamma, z) = f_\mathrm{LDDE}(L_\gamma, z=0) \frac{\dd{N}}{\dd{V}}(L_\gamma, z).
\end{equation}
In \citet{Ajello:2012kf, Ajello:2014lg}, the best-fit LDDE models for
BL Lacs and FSRQs both show that more luminous and rarer sources tend
to have density evolutions that peak at higher $z$, whereas the bulk
of lower-luminosity sources found at lower $z$. For FSRQs, the peak
redshift varies from $z_\mathrm{peak} \sim 1$--$2$ for sources with
$L_\gamma \sim 10^{46}$--$10^{49}$~erg~$\mathrm{s}^{-1}$. Similarly
for BL Lacs it varies from $z_\mathrm{peak} \sim 0$--$1$ for sources
with $L_\gamma \sim 10^{44}$--$10^{47}$~erg~$\mathrm{s}^{-1}$.

For the results shown in Sect.~\ref{sec:chance}, the driving factor is
the number of detected blazars and flares with which neutrino alert
events can be matched. The number of detected blazars in the redshift
range $z_\mathrm{min} < z < z_\mathrm{max}$ and above a $\gamma$-ray
flux limit, $F_\mathrm{lim}$, is given by
\begin{equation}
  N(>F_\mathrm{lim}) = \int_{z_\mathrm{min}}^{z_\mathrm{max}} \dd{z} \frac{ \dd{V} }{ \dd{z} } \int_{L_\mathrm{lim}(z)}^{\infty} \dd{L_\gamma} f(L_\gamma, z), 
\end{equation}
where $L_\mathrm{lim} = F_\mathrm{lim} 4 \pi D^2_L(z)$. For
$F_\mathrm{lim} \sim 10^{-12}$~erg~$\mathrm{cm}^{-2}~\mathrm{s}^{-1}$,
roughly at the detection threshold of the Fermi-LAT, we expect this
number to be dominated by lower-luminosity sources with similar
density evolutions for both the PDE and LDDE model and therefore
little impact on our results. Increasing $F_\mathrm{lim}$,
higher-luminosity sources will eventually begin to dominate in both
cases, but the highest luminosity sources are more likely to be found
at higher redshifts in the LDDE case and so will be observed with
lower fluxes. In this way, we could expect the chance coincidence
probability in the upper panel of
Fig.~\ref{fig:coincidence_prob_thresh} to fall off more steeply at
high $F_\gamma$.

For the results shown in Sect.~\ref{sec:connection}, the number of
neutrino alerts produced by the population is what drives the derived
constraints. For the assumptions on the blazar--neutrino connection
described in Sect.~\ref{sec:blazar_nu_connection} above, the LDDE
model could lead to a smaller contribution from the brightest sources,
leading to more relaxed constraints on $Y_{\nu\gamma}$.

However, in both Sects.~\ref{sec:chance} and \ref{sec:connection}, we
do not expect the impact of the LDDE model to have a greater effect
than that of the extreme `higher' and `lower' blazar population models
considered. The impact would at most be on the scale of the difference
between the BL Lac and FSRQ populations, which are treated similarly
to the low- and high-luminosity extremes of the LDDE model,
respectively. We can see the size of this effect clearly when the
results are shown independently of the variability of the two
populations. For example, in the top panel of
Fig.~\ref{fig:coincidence_dist}, or in the `All emission' case of
Figs.~\ref{fig:single_event_const} and
\ref{fig:total_event_const}. For our empirical flare model, the number
of flares and the flare properties are independent of $L_\gamma$, and
so we would not expect any further effects relevant for these results.

\end{document}